\documentclass[a4paper,11pt]{article}
\pdfoutput=1 
\usepackage{jheppub} 
\usepackage{bm}
\usepackage{cancel}
\usepackage[compat=1.1.0]{tikz-feynhand}
\usepackage[T1]{fontenc} 
\usepackage{slashed}
\newcommand{\ri}{{\rm i}}

\title{\boldmath More on scattering processes of dressed particles with a time-dependent mass }

\author[a]{Yusuke Yamada}

\affiliation[a]{Waseda Institute for Advanced Study, Waseda University, 1-21-1 Nishi Waseda, Shinjuku, Tokyo 169-0051, Japan}
\emailAdd{y-yamada@aoni.waseda.jp}

\abstract{We discuss the scattering process of a scalar field having a time-dependent mass with another scalar field having a constant mass as a toy model of the scattering problems during preheating after inflation. Despite a general difficulty of analytically solving such models, in our previous work~\cite{Taya:2022gzp}, we considered an exactly calculable model of such scattering processes with a time-dependent mass of the form $m^2(t)\supset \mu^4t^2$ and the time-dependence never disappears formally. In this work, we discuss another exactly calculable model with a time-dependent mass that has a spike/peak but asymptotes to a constant, which effectively appears in the preheating model of Higgs inflation with a non-minimal coupling.  Thanks to the localized time-dependence of the mass, the daughter particle number density behaves in a physically reasonable way contrary to the one in our previous model due to the infinite time-dependent mass in the asymptotic future. On the other hand, we find that the daughter particle experiences the kinematically forbidden process, which is a non-perturbative phenomenon found in our previous work. As in the previous model, the kinematically forbidden process produces daughter particles exponentially more than the parent particle having the time-dependent mass, which never happens for particle decay processes without time-dependent backgrounds. This result supports the existence of such a non-perturbative particle production process in general time-dependent backgrounds.}

\begin{document}
\maketitle
\flushbottom
\section{Introduction}
Quantum field theory under strong background fields have been studied in various contexts such as strong field quantum electrodynamics (QED), cosmology, blackhole spacetime. One of the most interesting phenomena in such a theory is the production of particles from vacuum, such as the Sauter-Schwinger effect~\cite{Sauter:1931zz,Schwinger:1951nm} in strong field QED and the Hawking radiation~\cite{Hawking:1975vcx} in black hole spacetime. In the context of cosmology, particles can be produced from vacuum e.g. by the expansion of the universe~\cite{Parker:1969au,Parker:1971pt} and by the coupling to inflaton field, which is called preheating~\cite{Dolgov:1989us,Traschen:1990sw,Kofman:1994rk,Shtanov:1994ce,Yoshimura:1995gc,Kofman:1997yn}. In each case, we are able to estimate the amount of the produced particle by solving the free quantum fields coupled to classical background fields. Although analytic solutions for such dynamics are not available in general, we are able to solve such dynamics numerically, particularly when the external classical fields depend only on a single coordinate such as time.

How is the energy of the particles produced from vacuum converted to other elements? Understanding of such dynamics is quite important to describe the thermal history of the universe after inflation. In literature, such processes are often described e.g. by the Boltmann equations based on the flat spacetime scattering processes. Fully quantum theoretical description of such processes has not been studied much due to the technical difficulties in computing such processes. Nevertheless, quantum field theoretical description including background fields can be important as there can be phenomena that cannot be captured by S-matrix computations within the quantum theory without background fields.

Indeed, nontrivial backgrounds lead to non-perturbative particle production processes within the scattering processes. In our previous work~\cite{Taya:2022gzp}, we studied a specific example of the decay of particles coupled to a strong field background on the basis of the fully quantum field theoretical calculations, which is the quantum field theoretical completion of instant preheating \cite{Felder:1998vq,Felder:1999pv}. It turns out that, in addition to the perturbative decay processes that can be understood from the standard perturbation theory, there appears a new particle production process that violates the energy conservation law of the quantum system under consideration, which we dub {\it kinematically forbidden processes}. Such processes result from the parent particles non-perturbatively dressed by the time-dependent background, and quantifying the amount of the produced daughter particles can be achieved only by the first-principle calculations, as far as we know. Depending on the model parameters, the kinematically forbidden processes can occur more efficiently than the perturbative decay processes, and therefore, such a process should be taken account of e.g. in cosmology.

In this work, we consider another calculable model of scatterings including a quantum field dressed by external classical background. There are several motivations to discuss a new model from both theoretical and phenomenological viewpoints. From a theoretical side, an analytically calculable model of such particle scattering process is quite few and is worth constructing if possible. One of the difficulties originates from the absence of the analytical solutions to free mode equation of quantum fields dressed by classical backgrounds when the background is complicated. Fully numerical calculations cost huge computational resources in addition to issues of numerical errors. In particular, non-perturbative particle productions are typically exponentially suppressed and small, which may be hidden in the possible errors in numerical simulations. In addition to the difficulty in finding analytical solutions to free equation of motion, scattering process requires to evaluate e.g. the time integration of products of mode functions of scattered particles, which is practically quite difficult to perform within numerical methods. 

From a phenomenological side, the model we consider in this work (with the time-dependent mass shown in Fig.~\ref{fig:mt}) is related to the preheating after Higgs inflation, where the non-minimal coupling of a Higgs field to Ricci scalar causes a spiky mass of particles when the Higgs field oscillates~\cite{Ema:2016dny,Sfakianakis:2018lzf,He:2018mgb}.\footnote{The time-dependent mass in Fig.~\ref{fig:mt} becomes spiky in the large $\mu$ limit where $\mu$ denotes a mass parameter in \eqref{mtdef}.} Our exact result can be thought of an approximation of such a situation.\footnote{The reheating process within the Higgs inflation is addressed e.g. in~\cite{He:2020qcb,Aoki:2022dzd} but the effect of the time-dependent mass is not fully taken account of in these works. Our work is complementary to them as ours uses a fully quantum field theoretical approach.} We will illustrate the relation between our toy model and the inflation models with non-minimal couplings in Sec.~\ref{nonminimalcoupling}. Although we will derive an exact result for the model with the time-dependent mass having a single peak, we also discuss how to generalize the result to the case with more peaks appear in the time-dependent mass, which approaches toward the preheating era in the Higgs inflation model. 

We emphasize the difference between our previous work~\cite{Taya:2022gzp} and the present one: In the model of~\cite{Taya:2022gzp}, we have considered a time-dependent mass $m^2(t)\sim \mu^4t^2$, where $\mu$ is a mass parameter, and therefore the mass of a dressed particle is time-dependent at any time and infinite for both asymptotic future and past at least at the formal level. As the parent particle energy becomes infinite for any momentum mode, the daughter particle number density in its phase space turns out to be non-convergent when integrated over its momentum. In this work, on the other hand, we consider a time-dependent mass whose time-dependence is localized at some time and decay exponentially both in the asymptotic past and future, which makes the particle picture of a dressed field clearer while leaving non-trivial time-dependence in some time region. As we will show explicitly, the daughter particle number density decays fast enough for a high momentum modes, such that the total number density in space can be finite. We also find a reasonable singularity associated with the standard perturbative decay process, which was absent in the previous model due to a continuously growing mass of a dressed field.

The rest of the paper is organized as follows: In Sec.~\ref{freefield}, we show the setup of the model and the exact solution to the free field equation of motion of the dressed scalar field, which is the basis of the following scattering problems. We then consider an interacting theory by introducing an additional scalar field having a time-independent mass in Sec.~\ref{Scattering}. We will show the behavior of the daughter particle number density (integrand) in various parameter limits of the exact result and briefly discuss how to generalize our result based on the time-dependent mass having a single peak to a more general situation with multiple peaks. In Sec.~\ref{nonminimalcoupling}, we give a few explanation how our toy model is related to inflationary models with non-minimal couplings to Ricci scalar. Finally we conclude in Sec.~\ref{conclusion}. The appendix~\ref{LFproperty} shows some mathematical properties of the associated Legendre functions and its integration that is used in the main text.

We will take the natural unit convention $\hbar=c=1$ and the metric is the mostly plus one $\eta_{\mu\nu}={\rm diag}(-1,+1,+1,+1)$.
\section{Free fields with time-dependent mass}\label{freefield}
\begin{figure}{t}
    \centering
    \includegraphics[width=0.5\linewidth]{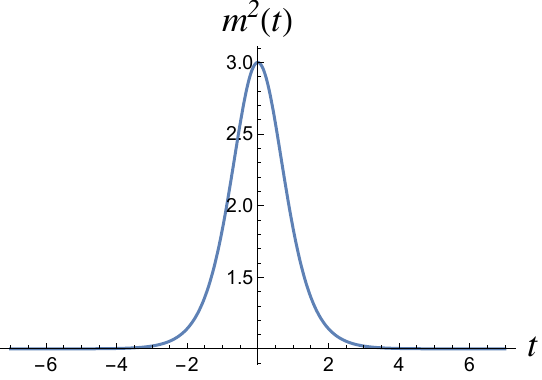}
    \caption{The time-dependent mass with $\mu=m_\phi=1$ and $\nu=1$.}
    \label{fig:mt}
\end{figure}

We consider a real scalar field $\phi$ having a time-dependent mass given by
\begin{align}
    m^2(t)=\frac{\nu(\nu+1)\mu^2}{\cosh^2 \mu t}+m_\phi^2,\label{mtdef}
\end{align}
where $\nu$ is a real positive parameter, $\mu$ is the mass parameter, and $m_\phi^2$ corresponds to the mass of $\phi$ in the asymptotic past and future. The above parametrization is just for convenience. We have shown the numerical example in Fig.~\ref{fig:mt}.
The inverse of the mass parameter $\mu$ characterizes the time scale of the time dependence while the size of the peak depend both on $\mu$ and $\nu$. The free action of $\phi$ is given by
\begin{align}
    S=-\frac12\int d^4x \left[(\partial\phi(x))^2+m^2(t)\phi^2(x)\right].
\end{align}
The corresponding free quantum field operator $\hat{\phi}^{(0)}(t,\bm x)$ is
\begin{align}
    \hat{\phi}^{(0)}(t,\bm x)=\int \frac{d^3{\bm k}}{(2\pi)^{\frac32}}\left[\hat{a}^{\rm in}_{\bm k} f_k(t)e^{\ri \bm k\cdot \bm x}+\hat{a}_{\bm k}^{{\rm in}\dagger} \bar{f}_k(t)e^{-\ri \bm k\cdot \bm x}\right],\label{qscalar}
\end{align}
where the creation and annihilation operators $\hat{a}_{\bm k}^{\rm in},\hat{a}_{\bm k}^{{\rm in}\dagger}$ satisfies $\left[\hat{a}^{\rm in}_{\bm k},\hat{a}^{{\rm in}\dagger}_{\bm k'}\right]=\delta^3(\bm k-\bm k')$,  $\bar{f}_k(t)$ is complex conjugate of $f_k(t)$, the mode function $f_k(t)$ satisfies
\begin{align}
    \ddot{f}_k(t)+\left(k^2+m^2+\frac{\nu(\nu+1)\mu^2}{\cosh^2 \mu t}\right)f_k(t)=0,\label{modeEq}
\end{align}
and a dot represents a time derivative $\frac{d}{dt}$. The mode equation is of the form of the one-dimensional Schr\"odinger equation with an inverted P\"oschl-Teller potential (see e.g. Chapter 3 of \cite{Landau:1991wop}). We have put an upper script $(0)$ on the quantum field to distinguish it with the operator evolved by an interaction Hamiltonian. The advantage of this toy model is that the exact solution to the mode equation is available, which is given by associated Legendre functions $P_\nu^\mu(x)$, $Q_\nu^\mu(x)$ of which definitions and properties are summarized in Appendix~\ref{LFproperty} (on the basis of \cite{2014776,2015867}). In particular, let us look for the adiabatic vacuum mode function $f_k(t)$ satisfying
\begin{align}
    f_k(t)\to\frac{1}{\sqrt{2\omega_k}}e^{-\ri\omega_k t}\qquad \text{as}\quad t\to-\infty,
\end{align}
where $\omega_k^2\equiv k^2+m_\phi^2$. Note that the mode equation implies the conservation of a combination $\dot{f}_k(t)\bar{f}_k(t)-\dot{\bar{f}}_k(t)f_k(t)$, and the above initial condition implies the normalization of the mode function
\begin{align}
\dot{f}_k(t)\bar{f}_k(t)-\dot{\bar{f}}_k(t)f_k(t)=-\ri.\label{fnormalization}
\end{align}
With such an asymptotic condition, the creation and annihilation operators $\hat{a}_{\bm k}^{{\rm in}\dagger}, \ \hat{a}_{\bm k}^{\rm in}$ can be identified as that for the asymptotic past $t\to-\infty$. We can define a state annihilated by $\hat{a}_{\bm k}^{\rm in}$ denoted by $|0\rangle_{\rm in}$, which is the vacuum state in the asymptotic past.  With a new ``time'' variable $\xi\equiv \tanh (\mu t)$, the past vacuum mode function is explicitly given by
\begin{align}
    f_k(\xi)=\frac{\Gamma(1-\ri r_k)}{\sqrt{2\omega_k}}\left\{\cosh(\pi(r_k-\ri\nu))P_\nu^{\ri r_k}(\xi)-\frac{2\ri}{\pi}\sinh(\pi(r_k-\ri\nu))Q_\nu^{\ri r_k}(\xi)\right\},\label{vacmodefunction}
\end{align}
where 
\begin{align}
    r_k\equiv\frac{\omega_k}{\mu}.
\end{align} On the other hand, the mode function that asymptotes to the negative frequency mode $\propto e^{+\ri\omega_k t}$ is given by a naive complex conjugation of $f^{-}_k(\xi)$:
\begin{align}
  \bar{f}_k(\xi)=\frac{\Gamma(1+\ri r_k)}{\sqrt{2\omega_k}}\left\{\cosh(\pi(r_k+\ri\nu))P_\nu^{-\ri r_k}(\xi)+\frac{2\ri}{\pi}\sinh(\pi(r_k+\ri\nu))Q_\nu^{-\ri r_k}(\xi)\right\},
\end{align}
where we have changed only the signs in front of the imaginary unit $\ri$. We note that the asymptotic expansion of the limit $t\to+\infty$ is
\begin{align}
    f_k(t)\to &\frac{\alpha_k}{\sqrt{2\omega_k}}e^{-\ri\omega_k t}+\frac{\beta_k}{\sqrt{2\omega_k}}e^{+\ri\omega_k t},\label{fmaysm}
\end{align}
where
\begin{align}
   \alpha_k=& \frac{ e^{\ri(\vartheta^\nu_p-\vartheta^{0}_p)}\sinh(\pi(r_k-\ri \nu))}{\sinh (\pi r_k)},\\
   \beta_k=&\frac{\ri\sin (\pi\nu)}{\sinh (\pi r_k)},
\end{align}
and $e^{\ri\vartheta^\nu_p}\equiv\frac{\Gamma(\nu+\ri r_p+1)}{\Gamma(\nu-\ri r_p+1)}$. These coefficients satisfy $|\alpha_k|^2-|\beta_k|^2=1$. The appearance of the positive frequency mode $\propto e^{+\ri\omega_k t}$ is due to the time-dependent background, and represents the particle production from vacuum: We may define the creation and annihilation operator as the coefficients of the positive and the negative frequency modes as
\begin{align}
    \hat{\phi}^{(0)}(t,\bm x)=\int \frac{d^3{\bm k}}{(2\pi)^{\frac32}}\left[\hat{a}^{\rm out}_{\bm k} \tilde{f}_k(t)e^{\ri \bm k\cdot \bm x}+\hat{a}_{\bm k}^{{\rm out} \dagger}\bar{\tilde{f}}_k(t)e^{-\ri \bm k\cdot \bm x}\right],
\end{align}
where $\tilde{f}_k(t)$ is a function satisfying the mode equation same as $f_k(t)$~\eqref{modeEq} but a different boundary condition $\tilde{f}_k(t)\to\frac{1}{\sqrt{2\omega_k}}e^{-\ri\omega_k t}$ as $t\to+\infty$. Thus-defined creation and annihilation operators can be regarded as that of asymptotic future particles. Taking the limit $t\to+\infty$ for \eqref{qscalar} (with the asymptotic form \eqref{fmaysm}), we find the relation between the future and the past ladder operators as
\begin{align}
    \hat{a}^{\rm out}_{\bm k}=&\alpha_k\hat{a}^{\rm in}_{\bm k}+\bar{\beta}_k\hat{a}^{{\rm in} \dagger}_{-\bm k},\\
    \hat{a}^{{\rm out}\dagger}_{\bm k}=&\bar{\alpha}_k\hat{a}^{{\rm in}\dagger}_{\bm k}+\beta_k\hat{a}^{\rm in}_{-\bm k},
\end{align}
and the expectation value of the number density operator (in the phase space) evaluated by the past vacuum is
\begin{align}
    \langle \hat{n}^{\phi \rm out}_{\bm k}\rangle_{\rm in}=\frac{{}_{\rm in}\langle 0|\hat{a}^{{\rm out}\dagger}_{\bm k}\hat{a}^{{\rm out}}_{\bm k}|0\rangle_{\rm in}}{\delta^3({\bm p}\to {\bm 0})}=|\beta_k|^2,
\end{align}
which can be interpreted as the particle number density produced from vacuum due to the time-dependent background. The classical time-dependent background breaks the time-translation symmetry of the quantum system under consideration, which leads to the violation of the energy conservation and accordingly the notion of a vacuum state is ambiguous unless the background is turned off. In our model, the time dependence of the mass is localized around the time $t=0$ within the time interval for $\mathcal{O}(\mu^{-1})$. The amount of the $\phi$-particle production is analytically expressed as
\begin{align}
    \langle\hat{n}_{\bm k}^{\phi \rm out}\rangle_{\rm in}=\frac{\sin^2(\pi\nu)}{\sinh^2\left(\frac{\pi\omega_k}{\mu}\right)}.\label{phiproduction}
\end{align}
One finds that the particle number density decays exponentially in $\omega_k$ for $\omega_k\gg \mu/\pi$ whereas it behaves as $\omega_k^{-2}$ for $\mu/\pi\gg \omega_k$. It is rather surprising that the peak size of the time-dependent mass controlled by the parameter $\nu$ affects the amount of the particle production in a sinusoidal form. In particular, however large the peak can be, the particle production can be small if $\nu\sim n\in\mathbb{Z}$. 

\section{Scattering of dressed particles and non-dressed particles}\label{Scattering}
In this section, we consider the scattering dynamics of the dressed quantum field $\hat{\phi}$. Here we introduce another scalar field $\chi(t,\bm x)$ having a time-independent mass and coupling to $\phi(t,\bm x)$ as
\begin{align}
    S_\chi=-\frac{1}{2}\int dt d^3\bm x\left[(\partial\chi(t,\bm x))^2+m^2_\chi\chi^2(t,\bm x)+\lambda\mu\phi(t,\bm x)\chi^2(t,\bm x)\right],
\end{align}
where we have normalized the dimensionful coupling between $\phi$ and $\chi$ by $\mu$ that characterize the time-scale of $m^2(t)$~\eqref{mtdef}, and then the dimensionless coupling $\lambda$ corresponds to the ratio between the dimensionful coupling and $\mu$, which is a useful parametrization for later discussion. The time-independence of the $\chi$-particle mass $m_\chi^2$ is assumed for a technical reason to derive analytic expressions of the particle production rate. 

Within the standard perturbation theory without nontrivial backgrounds, $\phi$-particle can decay into two $\chi$-particles if $m_\phi>2m_\chi$. In particular, since the time-dependence disappears roughly for $t>+\mu^{-1}$, such a decay process should take place away from the time $t=0$. Furthermore, as the mass of $\phi$ becomes large around $t=0$, it may be possible to produce heavy $\chi$. Such a possibility is closely related to the instant preheating scenario~\cite{Felder:1998vq,Felder:1999pv}, where a scalar field produced due to preheating decays into other particles when the effective mass becomes large. Furthermore, in our previous work~\cite{Taya:2022gzp}, it is shown that the violation of energy conservation due to explicit time-dependence of the background leads to the decay process where the parent particle can decay into daughter particles that have energy less than that of a parent particle, which is kinematically forbidden in the time-independent background by the energy conservation law. Note however that there is no a~priori reason to expect the kinematically forbidden particle production within the present model since the time-dependent mass considered here and in \cite{Taya:2022gzp} are quite different.

We will take the Furry picture perturbation theory~\cite{Furry:1951bef}, which is the interaction picture with the dressed mode function for the field coupled to the nontrivial background. In our model, $\hat\phi(t,\bm x)$ has the nontrivial mode function~\eqref{vacmodefunction} whereas the field $\chi(t,\bm x)$ has the standard plane wave mode function. The free quantum field $\hat{\chi}^{(0)}(t,\bm x)$ is expanded as
\begin{align}
    \hat{\chi}^{(0)}(t,\bm x)= \int \frac{d^3\bm k}{(2\pi)^{\frac32}}\left[\hat{b}^{\rm in}_{\bm k}g_k(t)e^{\ri \bm k\cdot\bm x}+\hat{b}^{{\rm in}\dagger}_{\bm k}\bar{g}_k(t)e^{-\ri \bm k\cdot\bm x}\right],\label{chiexpand}
\end{align}
where the creation and annihilation operators $\hat{b}_{\bm k},\hat{b}_{\bm k}^\dagger$ satisfy $[\hat{b}_{\bm k},\hat{b}_{\bm k'}^\dagger]=\delta^3(\bm k-\bm k')$ and the mode function is simply given by a plane wave $g_k(t)=\frac{1}{\sqrt{2\Omega_k}}e^{-\ri \Omega_k t}$ with $\Omega_k\equiv\sqrt{k^2+m_\chi^2}$ and $\bar{g}_k(t)$ is a complex conjugate of $g_k(t)$. The interaction Hamiltonian $\hat{H}_{\rm int}(t)$ is
\begin{align}
    \hat{H}_{\rm int}(t)=\frac12\lambda\mu\int d^3x\left[\hat{\phi}^{(0)}(t,\bm x)\left(\hat{\chi}^{(0)}(t,\bm x)\right)^2\right]
\end{align}
where each field is understood as that in \eqref{qscalar} and \eqref{chiexpand}. The exponentiated time-ordered and anti-time-ordered products of the interaction Hamiltonian yields the time-evolution operator $\hat{U}(t,-\infty)$ and its Hermitian conjugate respectively, and more explicitly
\begin{align}
    \hat{U}(t,-\infty)\equiv {\rm T}\exp\left[-\ri\int_{-\infty}^{t}\hat{H}_{\rm int}(t')dt'\right]=\hat{I}-\ri\int_{-\infty}^{t}\hat{H}_{\rm int}(t')dt'+\mathcal{O}(\lambda^2).
\end{align}
In the following discussion, we consider the correction to the quantum field operators at the first order in the coupling $\lambda$, and evaluate the produced particle number density of the order $\lambda^2$, which is effectively given by the square of the first order corrections.\footnote{One may wonder if the appearance of $\mathcal{O}(\lambda^2)$ requires to consider the contribution from the mixing of the second order corrections and the zero-th order effect. However, it turns out that due to the absence of the production of $\chi$-particle from vacuum, namely the zero-th order contribution, implies that the second order corrections in $\lambda$ to a field operator does not affect the final result of the produced particle number. Therefore, taking account only of the first order corrections for the field operator is enough.} Note that we assume that a tadpole of $\phi$ at $\mathcal{O}(\lambda)$ originating from the contraction of two $\chi$ is properly removed by a counter term, although it does not affect the particle number density of $\chi$ either.

Let us evaluate the time evolution of operators in the presence of the interactions.\footnote{The following discussion is the same as that e.g. in \cite{Otto:2016xpn,Taya:2022gzp}.} At the first order in $\lambda$, the field operator $\hat{\chi}(t,\bm x)$ is given by
\begin{align}
    \hat{\chi}(t,\bm x)=&\hat{U}^\dagger(t,-\infty)\hat{\chi}^{(0)}(t,\bm x)\hat{U}(t,-\infty)\nonumber\\
    =&\hat{\chi}^{(0)}(t,\bm x)-\ri \lambda\mu\int_{-\infty}^{t} dt'd^3{\bm y}[\hat{\chi}^{(0)}(t,\bm x),\hat{\chi}^{(0)}(t',\bm y)]\hat{\chi}^{(0)}(t',\bm y)\hat{\phi}^{(0)}(t',\bm y)+\mathcal{O}(\lambda^2).\label{chipert}
\end{align}
We are able to evaluate $\hat{\phi}$ in the same way but since we are interested in the produced number of $\chi$-particle, the above formula is sufficient for our purpose. Now, we need to define the creation and annihilation operators at the asymptotic future, which can be extracted by taking Klein-Gordon inner product of the field operator and asymptotic mode functions like the Lehmann–Symanzik–Zimmermann (LSZ) reduction~\cite{Lehmann:1954rq}:
\begin{align}
    \hat{b}^{\rm out}_{\bm k}=(+\ri)\lim_{t\to+\infty}\int d^3{\bm x}\frac{\bar{g}_k(t)e^{-\ri \bm k\cdot\bm x}}{(2\pi)^{3/2}}\overleftrightarrow{\partial_{t}}\hat{\chi}(t,\bm x),
\end{align}
where $f(t)\overleftrightarrow{\partial_{t}}g(t)=f(t)\dot{g}(t)-\dot{f}(t)g(t)$. With \eqref{chipert}, we obtain
\begin{align}
     \hat{b}^{\rm out}_{\bm k}= \hat{b}^{\rm in}_{\bm k}-\ri\lambda\mu\int^{\infty}_{-\infty}dt'd^3\bm y \frac{\bar{g}(t')e^{-\ri\bm k\cdot\bm y}}{(2\pi)^{\frac{3}{2}}}\hat{\chi}^{(0)}(t',\bm y)\hat{\phi}^{(0)}(t',\bm y).
\end{align}
The Hermitian conjugation yields $\hat{b}^{{\rm out}\dagger}$ and then we are able to evaluate the number density of (future) $\chi$-particle having spatial momentum $\bm k$ in the state $|0\rangle_{\rm in}$ as
\begin{align}
    \langle \hat{n}_{\bm k}^\chi\rangle_{\rm in}=\frac{{}_{\rm in}\langle0|\hat{b}^{{\rm out}\dagger}_{\bm k}\hat{b}^{\rm out}_{\bm k}|0\rangle_{\rm in}}{\delta^3({\bm p}\to {\bm 0})}
    =\lambda^2\int \frac{d^3\bm p}{(2\pi)^3}\left|\mu\int^{\infty}_{-\infty}dtg_{k}(t)g_{|\bm k-\bm p|}(t)f_{p}(t)\right|^2.\label{daughternumberformula}
\end{align}
Thus, the evaluation of the daughter particle number density (in phase space) boils down to the time integral of products of mode functions. Now, we define the integral as
\begin{align}
F_{\bm k,\bm p}=&\mu\int^{\infty}_{-\infty}dtg_{k}(t)g_{|\bm k-\bm p|}(t)f_{p}(t)\nonumber\\
=&\frac{1}{2\sqrt{\Omega_{k}\Omega_{|\bm k-\bm p|}}}\mu\int^{\infty}_{-\infty}dt e^{-\ri (\Omega_k+\Omega_{|\bm k-\bm p|})t}f_p(t)\nonumber\\
=&\frac{1}{2\sqrt{\mu\Omega_{k}\Omega_{|\bm k-\bm p|}}}\left[A_p\tilde{I}_{p,+}\left(\Omega_k+\Omega_{|\bm k-\bm p|}\right)+B_p\tilde{I}_{p,-}\left(\Omega_k+\Omega_{|\bm k-\bm p|}\right)\right],\label{Fkpdef}
\end{align}
where 
\begin{align}
    A_p\equiv&\frac{\Gamma(1-\ri r_p)}{\sqrt{2r_p}}\frac{\ri \sin\pi\nu}{\sinh\pi r_p}\label{Ap}\\
    B_p\equiv& \frac{\Gamma(1-\ri r_p)}{\sqrt{2r_p}}\frac{e^{\ri\vartheta^\nu_p}\sinh(\pi(r_p-\ri\nu))}{\sinh(\pi r_p)}\label{Bp},
\end{align}
 recalling  $e^{\ri\vartheta^\nu_p}=\frac{\Gamma(\nu+\ri r_p+1)}{\Gamma(\nu-\ri r_p+1)}$. The integral $\tilde{I}_{p,\pm}(\Omega)$ are defined in \eqref{IpmOdef} and we have analytically evaluated it in appendix~\ref{app:Integral}. The integral has an analytic expression as shown in \eqref{Ipmfinal}. Note that the coefficients $A_p,\ B_p$ can be read off from the past vacuum mode function expressed by $P^\mu_{\nu}(\xi)$ in \eqref{pastmode2}. 

Although we have evaluated the daughter particle number density analytically, the formula is quite complicated. Nevertheless, let us read off the physical information from the analytic formula. The produced particle number is formally written as
\begin{align}
    \langle \hat{n}_{\bm k}^\chi\rangle_{\rm in}=&\lambda^2\int \frac{d^3\bm p}{(2\pi)^3}|F_{\bm k,\bm p}|^2\nonumber\\
    =&\lambda^2\int \frac{d^3\bm p}{(2\pi)^3}\frac{1}{4\mu\Omega_{k}\Omega_{|\bm k-\bm p|}}\left|A_p\tilde{I}_{p,+}\left(\Omega_k+\Omega_{|\bm k-\bm p|}\right)+B_p\tilde{I}_{p,-}\left(\Omega_k+\Omega_{|\bm k-\bm p|}\right)\right|^2.\label{nchiformula}
\end{align}
Here, we show the explicit forms of integrands as
\begin{align}
  A_p\tilde{I}_{p,+}\left(\Omega_k+\Omega_{|\bm k-\bm p|}\right)
  =&\frac{\pi \sin(\pi\nu)\sqrt{r_p}\Gamma\left(-\frac{\ri}{2}\rho_{\bm k,\bm p}^{+} \right)}{\sqrt2\rho_{\bm k,\bm p}^{-}\sinh\left(\frac{\pi}{2}\rho_{\bm k,\bm p}^{-}\right)\sinh(\pi r_p)\Gamma\left(1+\nu\right)\Gamma\left(-\frac{\ri}{2}\rho_{\bm k,\bm p}^{+}-\nu\right)}\nonumber\\
    &\qquad\times{}_3F_2\left(-\ri r_p-\nu,-\nu-1,-\frac{\ri}{2}\rho_{\bm k,\bm p}^+;-\ri r_p,-\frac{\ri}{2}\rho_{\bm k,\bm p}^+-\nu;1\right),
\end{align} 
and
\begin{align}
    B_p\tilde{I}_{p,-}\left(\Omega_k+\Omega_{|\bm k-\bm p|}\right)=&\frac{\ri\pi\sqrt{r_p}e^{\ri\vartheta^\nu_p-\ri\vartheta^0_p}\sinh(\pi(r_p-\ri\nu))\Gamma\left(-\frac{\ri}{2}\rho_{\bm k,\bm p}^{-} \right)}{\sqrt2\rho_{\bm k,\bm p}^{+}\sinh\left(\frac{\pi}{2}\rho_{\bm k,\bm p}^{+}\right)\sinh(\pi r_p)\Gamma\left(1+\nu\right)\Gamma\left(-\frac{\ri}{2}\rho_{\bm k,\bm p}^{-}-\nu\right)}\nonumber\\
    &\qquad\times{}_3F_2\left(\ri r_p-\nu,-\nu-1,-\frac{\ri}{2}\rho_{\bm k,\bm p}^-;\ri r_p,-\frac{\ri}{2}\rho_{\bm k,\bm p}^{-}-\nu;1\right),
\end{align}
where $\rho_{\bm k,\bm p}^{\pm}=\tilde{\Omega}_k+\tilde{\Omega}_{|\bm k-\bm p|}\pm r_p$.

In the following, we will investigate the behavior of the $\chi$-particle number density integrand in the three different parameter regime:
\begin{itemize}
    \item Large $p$ limit and large $k$ limit
    \item Resonant limit $\rho_{\bm k,\bm p}^-\to 0$
    \item Kinematically forbidden process with $k,m_\chi\ll p<m_\phi$ where the perturbative ``decay'' is kinematically forbidden
\end{itemize}
The three regimes are special limits of parameters and a more general behavior requires to use the full expressions given above.
\subsection{Asymptotic behavior of number density formula}
We consider the asymptotic behavior of the integrand $F_{\bm k,\bm p}$ for various limit of parameters. First, in order to make sure the convergence of the $\bm p$-integral, let us take a limit $p\to\infty$, which implies $\rho_{\bm k,\bm p}^-\to \tilde{\Omega}_k- (k\cos\theta)/\mu=\delta\tilde{\Omega}_k$ where ${\bm k}\cdot\bm p=k p\cos \theta$ and $\rho_{\bm k,\bm p}^+\to 2p/\mu$. 
\begin{align}
  \left|A_p\tilde{I}_{p,+}\left(\Omega_k+\Omega_{|\bm k-\bm p|}\right)\right|\overset{p\to\infty}{\longrightarrow}&\ \frac{\pi \sin(\pi\nu)(p/\mu)^{\frac12+\nu}e^{-\pi p/\mu}}{\sqrt2\delta\tilde{\Omega}_k\sinh\left(\frac{\pi}{2}\delta\tilde{\Omega}_k\right)\Gamma\left(1+\nu\right)}\nonumber\\
    &\qquad\times\left|{}_3F_2\left(-\ri p/\mu-\nu,-\nu-1,-\ri p/\mu;-\ri p/\mu,-\ri p/\mu-\nu;1\right)\right|
\end{align}
and
\begin{align}
   \left| B_p\tilde{I}_{p,-}\left(\Omega_k+\Omega_{|\bm k-\bm p|}\right)\right|\overset{p\to\infty}{\longrightarrow}&\ \frac{\pi e^{-\pi p/\mu}\left|\Gamma\left(-\frac{\ri}{2}\delta\tilde{\Omega}_k\right)\right|}{4 \sqrt{p/\mu}\Gamma\left(1+\nu\right)\left|\Gamma\left(-\frac{\ri}{2}\delta\tilde{\Omega}_k-\nu\right)\right|}\nonumber\\
    &\qquad\times\left|{}_3F_2\left(\ri p/\mu-\nu,-\nu-1,-\frac{\ri}{2}\delta\tilde{\Omega}_k;\ri p/\mu,-\frac{\ri}{2}\delta\tilde{\Omega}_k-\nu;1\right)\right|.
\end{align}
We note that $\left|{}_3F_2\left(-\ri p/\mu-\nu,-\nu-1,-\ri p/\mu;-\ri p/\mu,-\ri p/\mu-\nu;1\right)\right|=0$ and then, we should take account of next order in $\mu/p$. Therefore, $F_{\bm k,\bm p}$ exponentially decays for $p\to \infty$, which ensures the convergence of the $\bm p$-integration that appears in the number density formula~\eqref{nchiformula}.

We next consider the limit $k\to\infty$ corresponding to the infinite energy modes of daughter particles $\chi$, which should also decay faster than any powers of $k$. In the limit $k\to\infty$, we find $\rho_{\bm k,\bm p}^\pm\to 2k/\mu$.
\begin{align}
     \left|A_p\tilde{I}_{p,+}\left(\Omega_k+\Omega_{|\bm k-\bm p|}\right)\right|
  \overset{k\to\infty}{\longrightarrow}&\ \frac{\pi \sin(\pi\nu)\sqrt{r_p}(k/\mu)^{\nu-1} e^{-\pi k/\mu}}{2\sqrt2 \sinh(\pi r_p)\Gamma\left(1+\nu\right)}\nonumber\\
    &\qquad\times\left|{}_3F_2\left(-\ri r_p-\nu,-\nu-1,-\ri k/\mu;-\ri r_p,-\ri k/\mu-\nu;1\right)\right|,
\end{align}
and
\begin{align}
    \left|B_p\tilde{I}_{p,-}\left(\Omega_k+\Omega_{|\bm k-\bm p|}\right)\right|\overset{k\to\infty}{\longrightarrow}&\ \frac{\pi\sqrt{r_p}|\sinh(\pi(r_p-\ri\nu))|(k/\mu)^{\nu-1}e^{-\pi k/\mu}}{2\sqrt2\sinh(\pi r_p)\Gamma\left(1+\nu\right)}\nonumber\\
    &\qquad\times\left|{}_3F_2\left(\ri r_p-\nu,-\nu-1,-\ri k/\mu;\ri r_p,-\ri k/\mu-\nu;1\right)\right|.
\end{align}
We have numerically checked that the hypergeometric function part does not increase and therefore, we have confirmed that $|F_{\bm k,\bm p}|$ decays exponentially also for $k\to\infty$, which ensures that the total number density of $\chi$-particle given by $\bm k$-integration of \eqref{nchiformula} is also convergent.

The behavior of daughter particle number density for large momentum $k/\mu\gg1$ is quite different from the model considered in our previous work~\cite{Taya:2022gzp} where the time-dependent mass increases as $\sim t^2$. There, the daughter particle number density decays only as $k^{-3}$. In such a case, the total particle number density given by integrating over $\bm k$ is log-divergent. This is a consequence of the unphysical time-dependence of the mass where $m^2(t)\to \infty$ for $t\to\pm \infty$. In such a case, all the parent particle modes having a particular momentum $\bm p$ acquires infinite energy and arbitrarily large momentum modes of $\chi$ can be created. On the other hand, in our present model, the time-dependent mass has maximum and production rate of daughter particle $\chi$ with large $k$ decays exponentially. Such an exponential decay behavior ensures the expectation value of operators having arbitrary but finite powers of derivative can be finite (except the usual UV divergences associated with virtual particles). 

It would also be worth mentioning that the number density of $\chi$ exponentially decays as a function of $k/\mu$. Then, for a large $\mu$ namely for a spiky time-dependent mass, daughter particle with a large momentum can be produced as naively expected from the decay process $\phi\to\chi\chi$. But we should emphasize that such a ``perturbative'' decay picture holds only for a particular parameter region discussed in the next subsection. Therefore, exponential decay of daughter particle density with the exponent $k/\mu$ is quite non-trivial.

\subsection{Resonance and perturbative decay}\label{pertdecay}
We show the relation between perturbative decay process and the non-perturbative formula~\eqref{nchiformula}. Although we have checked the convergence of the $\bm p$-integral for high energy modes, the integrand has singular configurations by which the integral is not convergent. More specifically, we define a surface ${\cal S}_{\bm k,\bm p}$ in $\bm p$-space on which
\begin{align}
    \omega_p=\Omega_k+\Omega_{|\bm k-\bm p|}\label{energycons}
\end{align}
or equivalently $\rho^-_{\bm k,\bm p}=0$. On the surface ${\cal S}_{\bm k,\bm p}$, the integrand $F_{\bm k,\bm p}$ diverges. The appearance of such a singularity is physically quite reasonable: Despite a time-dependent mass of $\phi$, its effect is localized around $t=0$, and apart from such a time region, $\phi$-particle can decay into $\chi$ via the standard perturbative decay process. In the absence of the time-dependent mass, time-translation invariance can be restored and then the energy conservation law should hold for any particle scattering processes. The condition \eqref{energycons} is nothing but the energy conservation law for the decay process $\phi\to \chi\chi$. In this sense, it is reasonable to have divergent integrand $F_{\bm k,\bm p}$ within such a parameter region like the derivation of the Fermi's golden rule. Equivalently, the divergent contribution corresponds to the contribution of the standard perturbative decay process.

Let us take a closer look at the behavior of $F_{\bm k,\bm p}$ near the surface $S_{\bm k,\bm p}$, and we find
\begin{align}
  A_p\tilde{I}_{p,+}\left(\Omega_k+\Omega_{|\bm k-\bm p|}\right)
  \overset{\rho_{\bm k,\bm p}^-\to 0}{\longrightarrow}&\ \frac{\sqrt{2r_p} \Gamma\left(-\ri r_p \right)}{\rho_{\bm k,\bm p}^{-}\Gamma\left(1+\nu\right)\Gamma\left(-\ri r_p-\nu\right)}\frac{\sin(\pi\nu)}{\sinh(\pi r_p)}\nonumber\\
    &\times\frac{d}{dx}\left\{{}_3F_2\left(\ri x-\ri r_p-\nu,-\nu-1,-\frac{\ri}{2}x-\ri r_p;-\ri r_p,-\frac{\ri}{2}x-\ri r_p-\nu;1\right)\right\}\biggr|_{x\to 0}.
\end{align} 
From the definition of the generalized hypergeometric function
\begin{align}
    &\frac{d}{dx}\left\{{}_3F_2\left(\ri x-\ri r_p-\nu,-\nu-1,-\frac{\ri}{2}x-\ri r_p;-\ri r_p,-\frac{\ri}{2}x-\ri r_p-\nu;1\right)\right\}\biggr|_{x\to 0}\nonumber\\
    =&\sum_{N=0}^\infty\frac{d}{dx}\left\{\frac{\Gamma(\ri x-\ri r_p-\nu+N)\Gamma(-\nu-1+N)\Gamma(-\frac{\ri}{2}x-\ri r_p+N)\Gamma(-\ri r_p)\Gamma(-\frac{\ri}{2}x-\ri r_p-\nu)}{\Gamma(\ri x-\ri r_p-\nu)\Gamma(-\nu-1)\Gamma(-\frac{\ri}{2}x-\ri r_p)\Gamma(-\ri r_p+N)\Gamma(-\frac{\ri}{2}x-\ri r_p-\nu+N)N!}\right\}\biggr|_{x\to 0}\nonumber\\
    =&-\frac{\ri \Gamma(1+\nu)}{2}\left(\frac{3\Gamma(-\ri r_p-\nu)}{\Gamma(1-\ri r_p)}-\frac{\Gamma(-\ri r_p)}{\Gamma(1-\ri r_p+\nu)}\right),
\end{align}
where we have performed the infinite sum by Mathematica, which leads to
\begin{align}
    A_p\tilde{I}_{p,+}\left(\Omega_k+\Omega_{|\bm k-\bm p|}\right)
  \overset{\rho_{\bm k,\bm p}^-\to 0}{\longrightarrow}&\ \frac{\sin(\pi\nu)}{\sqrt{2r_p}\rho_{\bm k,\bm p}^{-}\sinh(\pi r_p)}\left(3-\frac{e^{\ri\vartheta_p^\nu-\ri\vartheta^0_p}\sinh(\pi(r_p-\ri \nu))}{\sinh(\pi r_p)}\right).
\end{align}
We also find
\begin{align}
    B_p\tilde{I}_{p,-}\left(\Omega_k+\Omega_{|\bm k-\bm p|}\right) \overset{\rho_{\bm k,\bm p}^-\to 0}{\longrightarrow}&+\frac{\sqrt2 e^{\ri\vartheta^\nu_p-\ri\vartheta^0_p}\sinh(\pi(r_p-\ri\nu))}{\rho_{\bm k,\bm p}^{-}\sqrt{r_p}\sin(\pi\nu)\sinh^2(\pi r_p)}.\label{Bsing}
\end{align}
Here, we have assumed that $\nu\neq 1,2,\cdots$ and when $\nu=1,2,\cdots$ there appears no singular contribution at $\rho_{\bm k,\bm p}^-=0$ from $B_p\tilde{I}_{p,-}$ (see \eqref{Imlimintnu} for its explicit form). Thus, contributions from the surface ${\cal S}_{\bm k,\bm p}$ is summarized as\footnote{The integration on $S_{\bm k,\bm p}$ does not make sense as it simply diverges. However, as discussed below, by replacing the singular part with $\delta$-function, we find a reasonable result.}
\begin{align}
     \langle \hat{n}_{\bm k}^{\chi,\rm pert}\rangle_{\rm in}\approx&\int_{{\cal S}_{\bm k,\bm p}} \frac{d^3\bm p}{(2\pi)^3}\frac{(\lambda\mu)^2}{4\omega_p\Omega_{k}\Omega_{|\bm k-\bm p|}}\frac{\langle\hat{n}_{\bm p}^{\phi \rm out}\rangle_{\rm in}}{(\omega_p-\Omega_k-\Omega_{|\bm k-\bm p|})^2}\nonumber\\
     &\qquad\times\left|\frac{3}{\sqrt2}-\frac{e^{\ri\vartheta_p^\nu-\ri\vartheta^0_p}\sinh(\pi(r_p-\ri \nu))}{\sinh(\pi r_p)}\left(\frac{1}{\sqrt2}-\frac{\sqrt2 }{\sin^2(\pi\nu)}\right)\right|^2.
\end{align}
We are able to find the correspondence to the decay rate based on the standard perturbation theory, which reads the decay amplitude as
\begin{align}
    P(\phi\to\chi\chi)=&(\lambda\mu)^2\left|\int_{-\infty}^\infty dt \bar{f}^{\rm w.o.}_p(t)g_{k}(t)g_{|\bm k-\bm p|}(t)\right|^2\nonumber\\
    =&\left| \lambda\mu \frac{1}{2\sqrt{2\omega_p\Omega_{k}\Omega_{|\bm k-\bm p|}}}(2\pi)\delta(\omega_p-\Omega_{k}-\Omega_{|\bm k-\bm p|}) \right|^2\nonumber\\
    =&\frac{(\lambda\mu)^2}{8\omega_p\Omega_{k}\Omega_{|\bm k-\bm p|}}(2\pi)\delta(\omega_p-\Omega_{k}-\Omega_{|\bm k-\bm p|})T_{\rm tot},
\end{align}
where $\bar{f}^{\rm w.o.}_p(t)=\frac{1}{\sqrt{2\omega_p}}e^{+\ri \omega_p t}$ is the mode function of $\phi$ without a time dependent mass and $T_{\rm tot}=(2\pi)\delta(E\to 0)$ is the total time. We define the production rate of the unit time $\Gamma_{\phi\to\chi\chi}$ as
\begin{align}
    \Gamma_{\bm k,\bm p}^{\phi\to\chi\chi}=\frac{(\lambda\mu)^2}{8\omega_p\Omega_{k}\Omega_{|\bm k-\bm p|}},
\end{align}
with which we are able to rewrite the number density formula as
\begin{align}
    \langle \hat{n}_{\bm k}^{\chi,\rm pert}\rangle_{\rm in}\approx&\int_{{\cal S}_{\bm k,\bm p}} \frac{d^3\bm p}{(2\pi)^3}\frac{1}{(\omega_p-\Omega_k-\Omega_{|\bm k-\bm p|})^2}\Gamma_{\bm k,\bm p}^{\phi\to\chi\chi}\times 2\langle\hat{n}_{\bm p}^{\phi \rm out}\rangle_{\rm in}\nonumber\\
     &\qquad\times\left|\frac{3}{\sqrt2}-\frac{e^{\ri\vartheta_p^\nu-\ri\vartheta^0_p}\sinh(\pi(r_p-\ri \nu))}{\sinh(\pi r_p)}\left(\frac{1}{\sqrt2}-\frac{\sqrt2 }{\sin^2(\pi\nu)}\right)\right|^2.
\end{align}
If we replace the singular factor $(\omega_p-\Omega_k-\Omega_{|\bm k-\bm p|})^{-2}$ with $\left(\pi\delta(\omega_p-\Omega_k-\Omega_{|\bm k-\bm p|})\right)^2$ which is motivated by the relation
\begin{align}
    \frac{1}{(\omega_p-\Omega_k-\Omega_{|\bm k-\bm p|})\pm \ri\varepsilon}={\rm P.V.}\left(\frac{1}{\omega_p-\Omega_k-\Omega_{|\bm k-\bm p|}}\right)\mp \ri\pi \delta(\omega_p-\Omega_k-\Omega_{|\bm k-\bm p|}),
\end{align}
we obtain
\begin{align}
    \langle \hat{n}_{\bm k}^{\chi,\rm pert}\rangle_{\rm in}\to&\int_{{\cal S}_{\bm k,\bm p}} \frac{d^3\bm p}{(2\pi)^3}\Gamma_{\bm k,\bm p}^{\phi\to\chi\chi}\langle\hat{n}_{\bm p}^{\phi \rm out}\rangle_{\rm in}T_{\rm tot}(2\pi)\delta(\omega_p-\Omega_k-\Omega_{|\bm k-\bm p|})\nonumber\\
     &\qquad\times \frac12\left|\frac{3}{\sqrt2}-\frac{e^{\ri\vartheta_p^\nu-\ri\vartheta^0_p}\sinh(\pi(r_p-\ri \nu))}{\sinh(\pi r_p)}\left(\frac{1}{\sqrt2}-\frac{\sqrt2 }{\sin^2(\pi\nu)}\right)\right|^2.
\end{align}
Then, we find the $\chi$-production rate per unit time as
\begin{align}
    \Gamma^{\phi\to\chi\chi}_{\bm k}=&\int \frac{d^3\bm p}{(2\pi)^3}\Gamma_{\bm k,\bm p}^{\phi\to\chi\chi}\langle\hat{n}_{\bm p}^{\phi \rm out}\rangle_{\rm in}(2\pi)\delta(\omega_p-\Omega_k-\Omega_{|\bm k-\bm p|})\nonumber\\
     &\qquad\times \frac12\left|\frac{3}{\sqrt2}-\frac{e^{\ri\vartheta_p^\nu-\ri\vartheta^0_p}\sinh(\pi(r_p-\ri \nu))}{\sinh(\pi r_p)}\left(\frac{1}{\sqrt2}-\frac{\sqrt2 }{\sin^2(\pi\nu)}\right)\right|^2,\label{pertdecrate}
\end{align}
where we have omitted the region of the integral as it is restricted to ${\cal S}_{\bm k,\bm p}$ by the $\delta$-function. 

Let us interpret the derived formula for the $\chi$-production rate. Except the $\mathcal{O}(1)$ factor, the production rate is reasonable as it is given by the decay rate of $\phi\to\chi\chi$, $\Gamma_{\bm k,\bm p}^{\phi\to\chi\chi}$, multiplied by the number density of the parent particle $\phi$ produced from vacuum. Since the time-dependence of $\phi$-mass is localized around $t=0$ and for most of the time region the standard perturbative decay is possible, which causes the singularity of the $\chi$-number density formula for the configuration \eqref{energycons}. The time-dependence of the mass of $\phi$ does not vanish strictly, which would lead to the power type singularity for $\rho_{\bm k,\bm p}^{-}\to 0$ instead of $\delta$-function. If we treated the exponentially decaying mass of $\phi$ as perturbation, we would find $\delta$-function singularity. The appearance of nontrivial $\mathcal{O}(1)$ factor (in the second line of \eqref{pertdecrate}) should be a consequence of the time-dependent mass of $\phi$. Nevertheless, since the prescription used to derive the formula~\eqref{pertdecrate} (with respect to a singular factor) is ad-hoc, it is hard to find its physical meaning. We should emphasize that the singularity $\nu=1,2,\cdots$ is absent if we carefully treat the original formula. Therefore, we should not think of the increase of the rate around $\nu\to 1,2,\cdots$ as physical. (See the comments around \eqref{Bsing} and \eqref{Imlimintnu}.) Therefore, it is reasonable to take the factor to be $\mathcal{O}(1)$ despite a seemingly-possible large contribution for $\nu\approx 1,2,\cdots$.
Thus, we have found the relation between the singular contribution to the number density and the standard perturbative decay process $\phi\to \chi\chi$.

Although the integrand is divergent on the parameter configuration \eqref{energycons}, total number density of $\chi$ produced through this process should be bounded by the following reasons: We have found the correspondence to the perturbative decay. Then, the produced $\chi$-particle number density is not infinite, but should be understood to increase by the rate $\Gamma^{\phi\to\chi\chi}_{\bm k}$ per unit time. Then, the particular mode of a parent particle $\phi$ is converted to at most two $\chi$-particles. At most, all the parent particles allowing perturbative decay process are converted to $\chi$-particles, but once all $\phi$-particles are converted, there would be no further $\chi$-production.  Therefore, the total $\chi$-particle number density $n_\chi$ never exceeds $2n_\phi=2\int\frac{ d^3p}{(2\pi)^3}\langle\hat{n}_{\bm p}^{\phi \rm out}\rangle_{\rm in}$.

As we will show in the next subsection, the above consideration only applies to the daughter particle modes that can be produced by the perturbative decay process, and the non-perturbative process shown in the next subsection can realize $n_\chi\gg n_\phi$. 

\subsection{Kinematically forbidden processes}
In the previous subsection, we have looked at the resonant configuration~\eqref{energycons} where the energy conservation holds. In other words, in most of the parameter region, the energy conservation does not hold. Nevertheless, the quantity $F_{\bm k,\bm p}$ does not vanish, which never occurs in the absence of the time-dependent mass.

\begin{figure}
    \centering
    \includegraphics[width=0.8\linewidth]{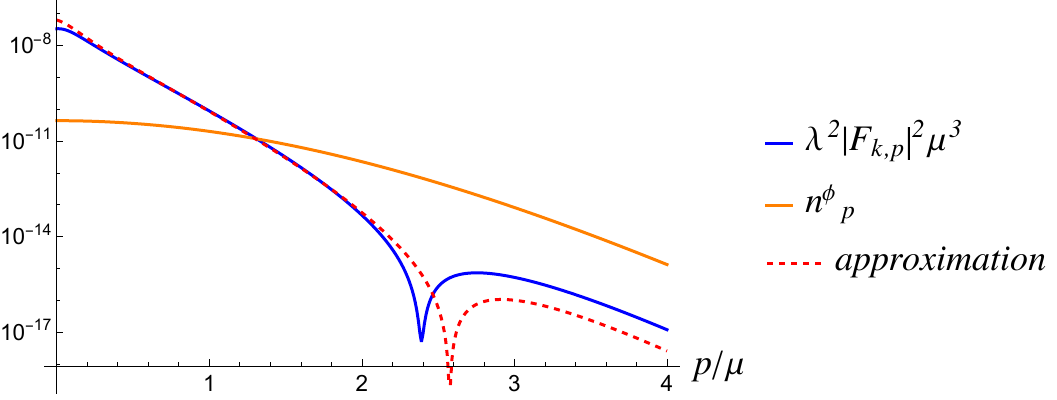}
    \caption{Comparison of the number density integrand $\lambda^2|F_{\bm k,\bm p}|^2\mu^3$ evaluated by the exact form, its approximation shown in~\eqref{KFpr}, and the number density of the parent particle $n_p^{\phi\rm out}$. We have chosen parameters $\nu=1.4, m_\phi=4\mu,m_\chi=0.1\mu,k=0.1\mu,\cos\theta=0.1, \lambda=0.1$.}
    \label{fig:kinematicallyforbidden}
\end{figure}

Among various parameter regions, we will focus on a particular situation where the daughter particle energy is much less than that of a parent particle, which forbids the perturbative decay process $\phi\to\chi\chi$. Furthermore, we consider the $\bm p$-integral region with $p\ll m_{\phi}$, namely
\begin{align}
   m_\chi,\  k\ll|\bm k-\bm p|, \ p< m_\phi.
\end{align}
In such a parameter region, \eqref{energycons} never holds and we may approximate $\rho_{\bm k,\bm p}^{\pm}\approx r_p\pm p/\mu$, which leads to
\begin{align}
 & A_p\tilde{I}_{p,+}\left(\Omega_k+\Omega_{|\bm k-\bm p|}\right)\nonumber\\
  \to &\frac{\pi \sin(\pi\nu)\sqrt{r_p}\left(\frac12(r_p+p/\mu)\right)^\nu e^{-\frac{\pi}{2}(3r_p-p/\mu)}}{\sqrt2(r_p-p/\mu)\Gamma\left(1+\nu\right)}\nonumber\\
    &\qquad\times{}_3F_2\left(-\ri r_p-\nu,-\nu-1,-\frac{\ri}{2}(r_p+p/\mu);-\ri r_p,-\frac{\ri}{2}(r_p+p/\mu)-\nu;1\right),
\end{align} 
and
\begin{align}
    &B_p\tilde{I}_{p,-}\left(\Omega_k+\Omega_{|\bm k-\bm p|}\right)\nonumber\\
    \to&\frac{\ri\pi\sqrt{2r_p}e^{\ri\vartheta^\nu_p-\ri\vartheta^0_p+\ri \pi\nu}\left(\frac12(p/\mu-r_p)\right)^\nu e^{-\frac{\pi}{2}(r_p+p/\mu)}}{(r_p+p/\mu)\Gamma\left(1+\nu\right)}\nonumber\\
    &\qquad\times{}_3F_2\left(\ri r_p-\nu,-\nu-1,-\frac{\ri}{2}(p/\mu-r_p);\ri r_p,-\frac{\ri}{2}(p/\mu-r_p)-\nu;1\right).
\end{align}
Then, within the parameter region under consideration $p<m_\phi$, the latter term dominates among the integrands, and we may approximate the $\chi$-particle density as
\begin{align}
    \langle \hat{n}_{\bm k}^\chi\rangle_{\rm in}\approx&\lambda^2\int_{p<m_\phi} \frac{d^3\bm p}{(2\pi)^3}\frac{1}{4\mu\Omega_{k}\Omega_{|\bm k-\bm p|}}\frac{2\pi^2r_p\left|\frac12(p/\mu-r_p)\right|^{2\nu} e^{-\pi(r_p+p/\mu)}}{(r_p+p/\mu)^2(\Gamma(\nu+1))^2}\nonumber\\
    &\quad\times\left|{}_3F_2\left(\ri r_p-\nu,-\nu-1,-\frac{\ri}{2}(p/\mu-r_p);\ri r_p,-\frac{\ri}{2}(p/\mu-r_p)-\nu;1\right)\right|^2 +\int_{p>m_\phi}(\cdots),\label{KFpr}
\end{align}
where the omitted part is contribution from larger momentum $p$ which is more suppressed exponentially.
Particularly from the small momentum region $p/\mu\ll m_\phi/\mu$, we find
\begin{align}
    \langle \hat{n}_{\bm k}^\chi\rangle_{\rm in}\approx&\lambda^2\int_{p\ll m_\phi} \frac{d^3\bm p}{(2\pi)^3}\frac{1}{4\mu\Omega_{k}\Omega_{|\bm k-\bm p|}}\frac{2\pi^2\left(\frac12r_0\right)^\nu e^{-\pi r_0 }}{\Gamma(\nu+1)}\nonumber\\
    &\quad\times\left|{}_3F_2\left(\ri r_0-\nu,-\nu-1,\frac{\ri}{2}r_0;\ri r_0,\frac{\ri}{2}r_0-\nu;1\right)\right|^2 +\cdots,\label{KFpr2}
\end{align}
This formula shows a remarkable property: Notice that the parent particle number density with small momentum is given by
\begin{align}
    \langle\hat{n}_{\bm 0}^{\phi \rm out}\rangle_{\rm in}\approx \sin^2(\pi\nu)e^{-2\pi r_0},
\end{align}
and compared with the daughter particle number density~\eqref{KFpr} particularly the contributions from $p/\mu\ll r_0= m_\phi/\mu$, the $\phi$-particle number density has an exponentially suppression factor with an exponent twice as large as that in $\chi$-particle number density formula. Equivalently, this fact implies that the daughter particle number density $n_\chi$ can be exponentially greater than that of the parent particle $n_\phi$. Such a situation cannot be realized by the standard perturbative decay process $\phi\to\chi\chi$, and the effect is non-perturbative. We have shown a numerical example of the daughter particle number density integrand and the parent particle number density $n_{\bm p}^{\phi\rm out}$ in Fig.~\ref{fig:kinematicallyforbidden}. Although the figure shows an integrand rather than integrated one, the resultant $\chi$-particle number would become exponentially larger than that of $\phi$ even after integrating over the small $p$ region. 

We emphasize that exponentially larger number density of the daughter particle than that of the parent particle is found also in our previous study \cite{Taya:2022gzp} where the time-dependent mass $\sim t^2$. Our observation supports that the kinematically forbidden particle production process is a rather general phenomenon and generally leads to exponentially larger amount of daughter particles than the parent particle.

One may wonder whether the kinematically forbidden process can produce more particles than through the perturbative decay process discussed in Sec.~\ref{pertdecay}. Although the particle number density (integrand) diverges for the configuration \eqref{energycons}, as explained in Sec.~\ref{pertdecay}, the divergence corresponds to the perturbative decay process $\phi\to\chi\chi$. From the process, a single $\phi$-particle produces two $\chi$-particles, and therefore the amount of $\chi$-particle produced from the perturbative decay process is at most twice as much as the number density of $\phi$. On the other hand, as we have shown above, the kinematically forbidden process produced $\chi$-particles exponentially more than $\phi$-particles, and therefore, for the case with a sufficiently large mass $r_0=m_\phi/\mu>1$ compared with $m_\chi/\mu$, the kinematically forbidden production dominates over the whole $\chi$-particle production processes.
\subsection{Note on the case with multiple particle production events}\label{multipleevent}

\begin{figure}
    \centering
    \includegraphics[width=0.7\linewidth]{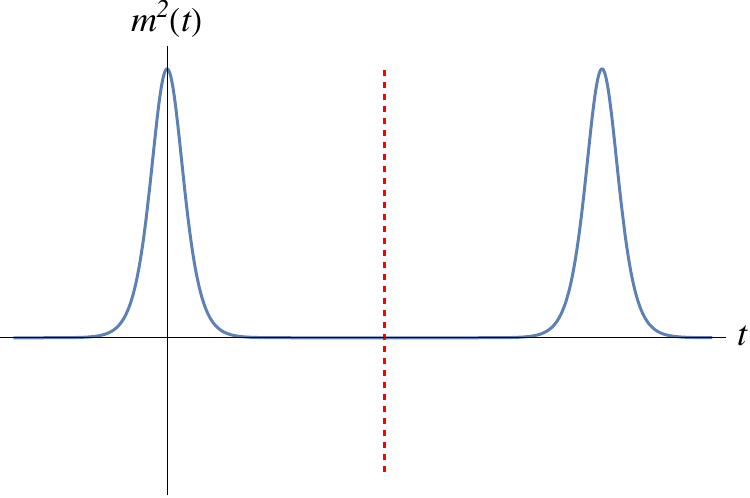}
    \caption{The time dependent mass with two peaks at different time. }
    \label{fig:multi-mass}
\end{figure}

In this subsection, we briefly discuss how to generalize our result to the case with a more general time-dependent mass. So far, we have discussed the case with the mass shown in Fig.~\ref{fig:mt}. One may wonder what would happen if the time-dependent mass has more peaks localized at different times as shown in Fig.~\ref{fig:multi-mass}. In order to approximate the case of Higgs inflation with a non-minimal coupling, we would need to add more peaks to the time-dependent mass. Note that the particle production with multiple peaks is studied also in~\cite{Amin:2015ftc} where the particle production associated with multiple peaks is treated stochastically. 

Although the time-dependent mass with two peaks seems a simple generalization, there is no exact analytic solution to the mode function. More generally, there appear multiple peaks with possibly different magnitude (or completely different time-dependence). Nevertheless, if the time-dependence effectively disappears for a sufficiently long time, we may consider the connection of exact solutions describing each time region. Specifically, let us consider a model with the time-dependent mass shown in Fig.~\ref{fig:multi-mass}. For the time region left to the red vertical line, we can approximate it by the exact past vacuum solution~\eqref{vacmodefunction}. If the two peaks are well separated, it would be possible to approximately describe the mode function of the time region right to the red vertical line with the functions appearing in~\eqref{vacmodefunction}. Nevertheless, to approximate the mode function in the second region, we need to consider connection of the mode function in the first region to the second region. From the asymptotic behavior of the past vacuum mode function~\eqref{fmaysm}, the appropriate ``positive frequency'' mode function describing the second region is
\begin{align}
    f_k^{\rm 2nd}(\xi)=\alpha_k f_k(\xi)+\beta_k \bar{f}_k(\xi),
\end{align}
where $f_k(\xi)$ is the past vacuum mode function~\eqref{vacmodefunction}. The asymptotic past limit of the above is
\begin{align}
    f_k^{\rm 2nd}\to\frac{1}{\sqrt{2\omega_k}}\left(\alpha_k e^{-\ri \omega_k t} +\beta_k e^{+\ri \omega_k t}\right) \quad \text{as }t\to-\infty,
\end{align}
which is continuation of~\eqref{fmaysm}.
Then, the asymptotic past limit of $f_k^{\rm 2nd}(\xi)$ reproduces \eqref{fmaysm}. The asymptotic future limit of $f_k^{\rm 2nd}(\xi)$ can be easily found as
\begin{align}
    f_k^{\rm 2nd}(\xi)\to \frac{\alpha_k^{\rm 2nd}}{\sqrt{2\omega_k}}e^{-\ri \omega_kt}+\frac{\beta_k^{\rm 2nd}}{\sqrt{2\omega_k}}e^{+\ri \omega_kt} \quad \text{as }t\to+\infty,
\end{align}
where 
\begin{align}
    \alpha_k^{\rm 2nd}=&\alpha_k^2+|\beta_k|^2,\\
    \beta_k^{\rm 2nd}=&\beta_k(\alpha_k+\bar{\alpha}_k).
\end{align}
We note that it is easy to check that $|\alpha_k^{\rm 2nd}|^2-|\beta_k^{\rm 2nd}|^2=1$ and the normalization condition is satisfied. Furthermore, the amount of the parent particle $\phi$ produced from vacuum after the secondary mass peak can be evaluated as
\begin{align}
    \langle\hat{n}_{\bm k}^{\phi \rm out}\rangle_{\rm in}^{\rm 2nd}=\left|\beta_k^{\rm 2nd}\right|^2
\end{align}

\begin{figure}
    \centering
    \includegraphics[width=0.8\linewidth]{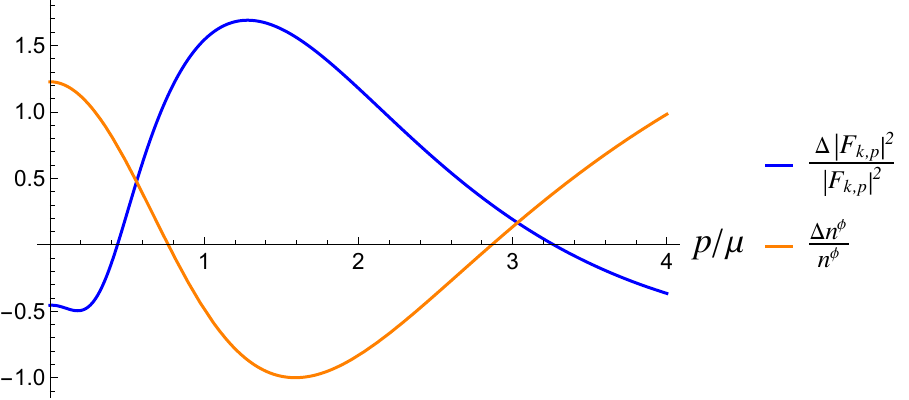}
    \caption{The difference of the particle number density within the model having time-dependent mass with one or two peaks. The horizontal axis denotes the momentum of the parent particle. We have chosen parameters $\nu=1.4, m_\phi=\mu,m_\chi=0.1\mu,k=0.1\mu,\cos\theta=0.1, \lambda=0.1$, where only $m_\phi$ differs from the ones in \ref{fig:kinematicallyforbidden}.}
    \label{fig:multi-ratio}
\end{figure}

If one is interested in the scattering event after the second peak, one may use $f_k^{\rm 2nd}(\xi)$ to evaluate the number density of daughter particle $\chi$. Since the mode function $f_k^{\rm 2nd}(\xi)$ is yet a linear combination of associated Legendre functions, we may use the exact result~\eqref{nchiformula}, but we should replace the constant coefficients $A_p,B_p$ by
\begin{align}
    \tilde{A}_p=&\alpha_p A_p+\beta_p\bar{B}_p,\\
    \tilde{B}_p=&\alpha_p B_p+\beta_p \bar{A}_p,
\end{align}
where $A_p, B_p$ are defined in \eqref{Ap} and \eqref{Bp}. We numerically evaluate how the secondary peak of the time-dependent mass of the parent particle $\phi$ affects the number density of both the parent $\phi$- and daughter $\chi$-particle number density (integrand) in Fig.~\ref{fig:multi-ratio}. In the figure, we have shown the ratio defined as
\begin{align}
    \frac{\Delta n^\phi_p}{n_\phi}\equiv& \frac{\left\langle\hat{n}_{\bm p}^{\phi \rm out}\right\rangle^{\rm 2nd}_{\rm in}-\left\langle\hat{n}_{\bm p}^{\phi \rm out}\right\rangle_{\rm in}}{\left\langle\hat{n}_{\bm p}^{\phi \rm out}\right\rangle_{\rm in}},\\
    \frac{\Delta F_{\bm k,\bm p}}{F_{\bm k,\bm p}}\equiv&\frac{\left|F_{\bm k,\bm p}^{\rm 2nd}\right|^2-\left|F_{\bm k,\bm p}\right|^2}{\left|F_{\bm k,\bm p}\right|^2},
\end{align}
where $F_{\bm k,\bm p}^{\rm 2nd}$ is the one \eqref{Fkpdef} with $A_p,B_p\to \tilde{A}_p,\tilde{B}_p$ defined above. Note that the figure should not be interpreted as that the decrease of the parent particle density is due to more production of daughter particle density (integrand) since the amount of the parent particle density $\langle\hat{n}_{\bm k}^{\phi \rm out}\rangle^{\rm 2nd}_{\rm in}$ is the one produced from vacuum and the effect of the $\chi$-particle production is not taken account of.

The evaluation of the scattering or decay of particles by connecting the mode functions in different time region would be a rather general method which can be applicable to other models. We note that both $\langle\hat{n}_{\bm p}^{\phi \rm out}\rangle^{\rm 2nd}_{\rm in}$ and $|F_{\bm k,\bm p}^{\rm 2nd}|^2$ exponentially decay for $p\to\infty$ (or $k\to \infty$), and hence the convergence property of the particle number density holds even for the secondary region as it should. 

We finally note that when there appear multiple peaks, we expect the kinematically forbidden process would occur at each peak, although it is difficult to prove it because of the lack of complete analytic estimation beyond the above mentioned approximation. As is clear, the kinematically forbidden process is caused by the time-dependence of the mass, and therefore, it occurs around the peak. If the multiple peaks are well separated in time, each peak can be thought of an independent event and the kinematically forbidden process would occur at each. Therefore, in the presence of multiple peaks, the kinematically forbidden processes would produce large amount of daughter particles. Such a possibility cannot be captured by the standard perturbative estimation based on the standard S-matrix and e.g. the Boltzmann equations.

\section{Relation to inflation with a non-minimal coupling}\label{nonminimalcoupling}
\begin{figure}
    \centering
    \includegraphics[width=0.7\linewidth]{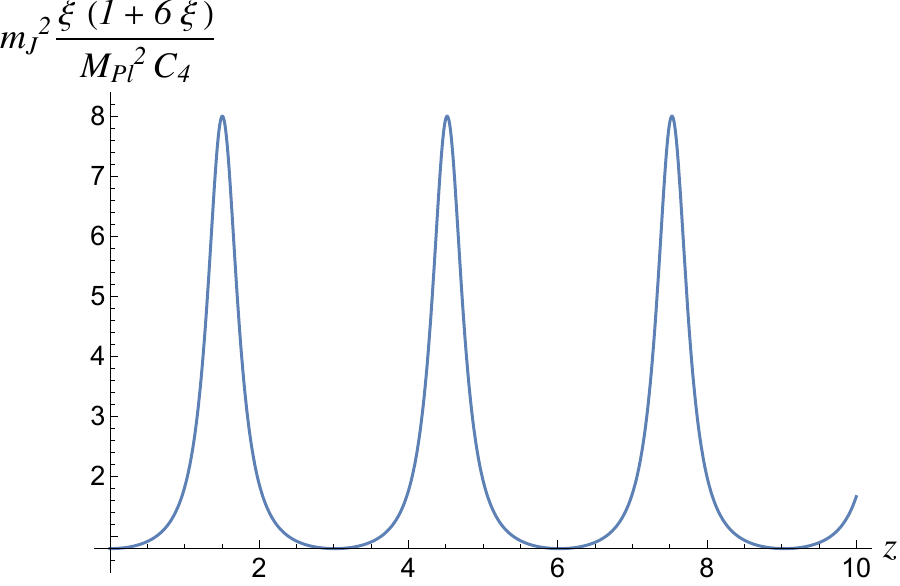}
    \caption{The behavior of $m_J^2$~\eqref{mJsq} with the numerical solution to \eqref{approxphi}. Here we have used the initial condition $\tilde{h}_J(z=0)=2,\frac{d}{dz}\tilde{h}_J(0)=0$. We have chosen these parameters just for an illustrating purpose. A larger initial value of $\tilde{h}_J(0)$ leads to sharper peaks. Note also that the vertical axis is dimensionless or equivalently, this shows the effective mass $m_J$ as the unit of a new mass scale $\sim \sqrt{C_4}M_{\rm Pl}/\xi\sim 4.8\times 10^{13}{\rm GeV}$ for the value of $\xi$ consistent with cosmic microwave background observations~\cite{Planck:2018jri}. Notice also that $z$ is the Jordan frame time in the unit of the same mass scale. }
    \label{fig:spikemass}
\end{figure}

On the basis of~\cite{Ema:2016dny}, we briefly illustrate the relation of our toy model to inflationary models with a real scalar inflaton field $h_J$ non-minimally coupled to Ricci scalar in which the action is
\begin{align}
    S=\int d^4x_J\sqrt{-g_J}\left[\left(\frac{M_{\rm Pl}^2}{2}+\frac{\xi}{2}h_J^2\right)R_J-\frac{1}{2}\partial_\mu h_J g_J^{\mu \nu}\partial_\nu h_J-V_J(h_J)\right],
\end{align}
where $x_J^\mu$ denotes the Jordan frame coordinates, $M_{\rm Pl}\sim 2.4\times10^{18}$ GeV is the reduced Planck scale, $\xi(\gg1)$ a coupling constant, $R_J$ the Ricci scalar (in Jordan frame), and $V_J(h_J)$ denotes the scalar potential of $h_J$ in the Jordan frame. We choose a quartic potential motivated by the Higgs field in standard model as $V_J(h_J)=\frac14C_4h_J^4$ with a coupling constant $C_4$. The Jordan frame metric is chosen as $ds^2=-dt_J^2+a^2_J(t_J)d\bm x_J^2$ where $a_J(t_J)$ denotes the scale factor. The equation of motion of $h_J$ in the Jordan frame becomes
\begin{align}
&3M_{\rm Pl}^2H_J^2+\xi\left(3H_J^2h_J^2+6H_Jh_J\frac{dh_J}{dt_J}\right)=\frac12\left(\frac{dh_J}{dt_J}\right)^2+\frac14C_4h_J^4\\
&(M_{\rm Pl}^2+\xi h_J^2)\left(2\frac{d H_J}{dt_J}+3H_J^2\right)+2\xi \left[h_J\frac{d^2h_J}{dt_J^2}+\left(\frac{dh_J}{dt_J}\right)^2+2H_Jh_J\frac{dh_J}{dt_J}\right]=-\frac12\left(\frac{dh_J}{dt_J}\right)^2+\frac14C_4h_J^4\\
    &\frac{d^2}{dt_J^2}h_J+3H_J \frac{d}{dt_J}h_J+m_J^2 h_J=0,
\end{align}
where $H_J\equiv a_J^{-1}\frac{d a_J}{dt_J}$ and
\begin{align}
    m_J^2\equiv\frac{\xi(1+6\xi)\left(\frac{dh_J}{dt_J}\right)^2+C_4M_{\rm Pl}^2h_J^2}{M_{\rm Pl}^2+\xi(1+6\xi)h_J^2}.\label{mJsq}
\end{align}
Note that the first two are derived by the variation of gravitational fields, whereas the third one is the equation of motion of $h_J$ rewritten by the use of the first two equations. In particular, the third equation just after the end of the inflation, where the inflaton oscillation amplitude is about $h_J^{\rm amp}\approx M_{\rm Pl}/\sqrt{\xi}$, can be approximately given by (neglecting the Hubble friction term)
\begin{align}
   \frac{ d^2\tilde{h}_J}{dz^2}+\frac{\left(\frac{d\tilde{h}_J}{dz}\right)^2+\tilde{h}_J^2}{1+\tilde{h}_J^2}\approx 0,\label{approxphi}
\end{align}
where $z=\frac{\sqrt{C_4}}{\sqrt{\xi(1+6\xi)}}M_{\rm Pl}t_J$ is a dimensionless time and $\tilde{h}_J=\frac{\sqrt{\xi(1+6\xi)}}{M_{\rm Pl}}h_J$ is also a dimensionless field. We can check that when initial amplitude $\tilde{h}_J$ is large enough, $\frac{d\tilde{h}_J}{dz}$ sharply changes in time. We have numerically solved the approximated equation of motion \eqref{approxphi} and shown the behavior of the effective mass $m_J^2$ as a function of $z$ in Fig.~\ref{fig:spikemass}. Up to the constant overall factor, the figure shows that the effective mass $m_J^2(z)$ resembles the time-dependent mass that we have used (in Figs.~\ref{fig:mt}, \ref{fig:multi-mass}). The sharpness of the peaks are related to the mass parameter $\mu$ in \eqref{mtdef}. For the Higgs inflation case, the mass scale characterizing the sharpness of the peak denoted as $m_{\rm sp}$ in \cite{Ema:2016dny} becomes $\mu\sim m_{\rm sp}=\frac{\sqrt{C_4}\xi (h^{\rm amp}_J)^2}{M_{\rm Pl}}$ where $h^{\rm amp}_J$ is the amplitude of the oscillation of $h_J$. Just after the end of inflation, $\mu\sim m_{\rm sp}\to \sqrt{C_4} M_{\rm Pl}$ which implies that rather heavy fields can be created during the preheating phase due to the spiky mass. The amplitude in the figure does not decay but if we take account of the Hubble friction, the amplitude of $h_J$ and accordingly the effective mass decreases in time.

We consider another massless scalar field $\phi_J$ that has the kinetic term of the form
\begin{align}
    S_\phi=\int d^4x_J\sqrt{-g_J}\left[-\frac12h_J^2\partial_\mu\phi_Jg^{\mu\nu}_J\partial_\nu\phi_J\right].\label{phikin}
\end{align}
Such a nontrivial kinetic term naturally appear if $h_J$ and $\phi_J$ is packaged into a complex scalar $\Phi=h_J e^{\ri \phi_J}$. After Weyl transformation $g_{\mu\nu}=\Omega g_{J\mu\nu}$ and field redefinition, we find
\begin{align}
    S_\phi=\int d\tau d^3\bm x\left[\frac12(\phi')^2-\frac12(\partial_i\phi)^2-\frac12m_\phi^2\phi^2\right]
\end{align}
where the Einstein frame metric is taken to be $ds^2=a^2(\tau)(-d\tau^2+d\bm x^2)$ and the prime denotes the derivative with respect to the conformal time $\tau$. The effective mass is given by
\begin{align}
    m_\phi^2=-\frac{(ah_J/\Omega)''}{(ah_J/\Omega)}\supset \frac{a^2 m_J^2}{\Omega^2}.
\end{align}
Note that since the spike appears near $\phi_J\sim 0$, the Jordan frame function $\Omega$ does not much affect the shape of the effective mass particularly around the peaks. Then, the effective mass of $\phi$ is almost the same as $m_J^2$. Since the shape of $m_J^2$ resembles the time-dependent mass, the $\phi$ field considered above can be approximately described by the one discussed in the previous section. We also note that the difference of the Jordan frame time and the Einstein frame time, or the difference in the conformal time and the standard time does not change the shape of the effective mass simply because the scale factor is slowly varying and $\Omega$ is not so large compared with the spike scale $m_{\rm sp}\sim \mu$. 

If the inflaton $h_J$ is identified as the Higgs field in the standard model, the phase mode $\phi_J$ is absorbed into a longitudinal mode of W-boson.\footnote{More precisely, Higgs scalar field is a doublet under SU(2) and three of them (as real scalars) are absorbed into W-and Z-boson. } Even in such a case, the effective mass of the longitudinal mode is of the same form at the leading order. Then, the daughter particles would be the standard model particles or weakly interacting massive particles (WIMP). We should note that the above mechanism is not restricted to Higgs inflation but is applied to the inflationary model with a non-minimal coupling. Furthermore, the field $\phi$ is not necessarily a phase direction of a complex scalar field. The nontrivial kinetic couplings between scalar fields and also the non-minimal coupling to gravity are ubiquitous e.g. in supergravity models and if there appear the coupling like the one in \eqref{phikin}, the effective time-dependent mass appear in the similar way, which causes production of $\phi$ and accordingly the scatterings with daughter particles described in this work. There appear multiple possibilities to generalize our model to more realistic inflation models, and therefore, we do not discuss any specific realization, but believe that our result does not change much by the  reason mentioned next.

We finally note that, although we have considered the parent and the daughter particles to be scalars, it is not essential: Since the most technically involved part is to solve the mode equation and to integrate the products of mode functions. As long as the mode function follows the similar Klein-Gordon equation with a (similar) time-dependent mass, only some kinematical factor differs from the case of scalars, but most of the computations such as the integration including the associated Legendre functions that we have derived should be applicable to such cases. Thus, our results regarding the scattering processes of dressed particles can be applied to models including particles with other spins. 

\section{Conclusion}\label{conclusion}
In this work, we have investigated scattering processes of a particle $\phi$ having a time-dependent mass~\eqref{mtdef} with $\chi$-particles having a constant mass, particularly the $\chi$-particle number density, which can be evaluated analytically. Unlike the previously considered model, the time-dependent mass in the present model is finite and its time-dependence is localized, which lead to the finite (spatial) number density of $\chi$, or equivalently, the fast decay of the number density for a large momentum limit of a daughter particle $\chi$. We have found the singular behavior for a resonant parameter configuration~\eqref{energycons}, which corresponds to the standard perturbative decay process $\phi\to\chi\chi$. Such a process should exist since the time-dependence disappears away from $t=0$.  We have also found the kinematically forbidden process where the kinematic condition~\eqref{energycons} does not hold and therefore is forbidden in the standard perturbation theory. As is the case of the previous model~\cite{Taya:2022gzp}, via the kinematically forbidden processes, light daughter particles can be produced exponentially more than the parent particle $\phi$. We have also considered the generalization to the case with multiple peaks in the time-dependent mass by matching the asymptotic forms of exact solutions.

As the kinematically forbidden particle production turns out to occur in two models (in \cite{Taya:2022gzp} and in the present work) with a rather different time-dependent mass, such a process would be a rather general phenomenon within scattering processes in nontrivial time-dependent backgrounds. It would be worth looking for other exactly calculable models with nontrivial time-dependent masses or possibly spacetime curvatures in order to check the presence of such non-perturbative processes. 

However, as quoted in the introduction, there is no exact solution for free fields within time-dependent backgrounds in general, and it is generally difficult to calculate the scattering rates or the amount of produced particles even if exact solutions to mode equations are available. In order to overcome such a difficulty, it would be important to develop approximation methods. One of the possibilities is to use the Wentzel-Kramers-Brillouin (WKB) approximation to mode functions. In particular, recent investigations revealed that the (free) particle production from vacuum in nontrivial time-dependent backgrounds can be quantitatively evaluated by investigating Stokes phenomena of mode equations~\cite{Dumlu:2010ua,Dumlu:2010vv,Dumlu:2011cc,Dumlu:2011rr,Enomoto:2020xlf,Taya:2020dco,Hashiba:2021npn,Yamada:2021kqw,Hashiba:2022bzi}. Although the particle production from vacuum is systematically understood as Stokes phenomena, it has not been achieved to evaluate scattering processes of dressed particles by using the WKB solutions, which would be worth developing also for computations of higher-order corrections such as loops.

It is also important to apply our results e.g. to cosmological models particularly for the preheating stage after inflation. In particular, the kinematically forbidden processes would produce relatively light species exponentially more than the parent particles. Taking account of such phenomena may change the thermal histories in known models. We will address applications of our result to realistic models elsewhere.

\section*{Acknowledgement}
This work is supported by Waseda University Grant for Special Research Projects (Project number: 2024C-773). The author would like to thank Hiroyuki Abe for useful discussions.

\appendix
\section{Properties of associated Legendre functions}\label{LFproperty}
In this appendix, we summarize formulas of associated Legendre functions from \cite{2014776,2015867}, which will be often used in the main text. The associated Legendre functions $P_\nu^\mu(\xi)$, $Q_\nu^\mu(\xi)$ ($\mu,\nu\in \mathbb{C}$) satisfy
\begin{align}
(1-\xi^2)\frac{d^2P_\nu^\mu(\xi)}{d\xi^2}-2\xi\frac{d P_\nu^\mu(\xi)}{d\xi}+\left[\nu(\nu+1)-\frac{\mu^2}{1-\xi^2}\right]P_\nu^\mu(\xi)=0,
\end{align}
and the same for $Q_\nu^\mu(\xi)$.
In our context, the value of $\xi$ is restricted to the real interval $\xi\in[-1,1]$, we are allowed to use the relation
\begin{align}
    P_\nu^\mu(\xi)=&\frac{1}{\Gamma(1-\mu)}\left(\frac{1+\xi}{1-\xi}\right)^{\frac{\mu}{2}}F\left(-\nu,\nu+1;1-\mu;\frac{1-\xi}{2}\right),\label{PFtrans}\\
    Q_\nu^\mu(\xi)=&\frac{\pi}{2\sin(\pi\mu)}\left[\cos(\pi\mu)P^\mu_\nu(\xi)-\frac{\Gamma(\nu+\mu+1)}{\Gamma(\nu-\mu+1)}P_{\nu}^{-\mu}(\xi)\right],\label{QPtrans}
\end{align}
where $\Gamma(z)$ is the $\Gamma$-function and $F(\alpha,\beta;\gamma;z)$ is the hypergeometric function.

The above identity for $Q^\mu_\nu(\xi)$ allows us to rewrite a part of the past vacuum mode function~\eqref{vacmodefunction} as
\begin{align}
    f^{-}_k(\xi)\supset&-\frac{\Gamma(1-\ri r_k)}{\sqrt{2\omega_k}}\frac{2\ri}{\pi}\sinh(\pi(r_k-\ri\nu))Q_\nu^{\ri r_k}(\xi)\nonumber\\
    =&-\frac{\Gamma(1-\ri r_k)\sinh(\pi(r_k-\ri\nu))}{\sqrt{2\omega_k}\sinh(\pi r_k)}\left[\cosh(\pi r_k)P^{\ri r_k}_\nu(\xi)-e^{\ri\vartheta^\nu_k}P^{-\ri r_k}_{\nu}(\xi)\right]\nonumber\\
    =&-\frac{\Gamma(1-\ri r_k)}{\sqrt{2\omega_k}}\left[\frac{\sinh(\pi(r_k-\ri\nu))}{\tanh(\pi r_k)}P^{\ri r_k}_\nu(\xi)-\frac{e^{\ri\vartheta^\nu_k}\sinh(\pi(r_k-\ri\nu))}{\sinh(\pi r_k)}P^{-\ri r_k}_{\nu}(\xi)\right].
\end{align}
 Thus, the past vacuum mode function can be rewritten as
\begin{align}
    f^{-}_k(\xi)=\frac{\Gamma(1-\ri r_k)}{\sqrt{2\omega_k}}\Biggl[&\frac{\ri \sin\pi\nu}{\sinh\pi r_k}P_\nu^{\ri r_k}(\xi)+\frac{e^{\ri\vartheta^\nu_k}\sinh(\pi(r_k-\ri\nu))}{\sinh\pi r_k}P^{-\ri r_k}_{\nu}(\xi)\Biggr].\label{pastmode2}
\end{align}
\subsection{Integral for production rate}\label{app:Integral}
In the main text, we are concerned with the mode function of daughter particles perturbed by the interaction with the dressed parent particle. In general, we need to perform integration of the form,
\begin{align}
    I_k(\Omega)=&\mu\int^{+\infty}_{-\infty}dt e^{-\ri\Omega t}f_k(t)\nonumber\\
    =&\int^{+1}_{-1}d\xi (1-\xi^2)^{-1}(1-\xi)^{\frac{\ri\tilde{\Omega}}{2}}(1+\xi)^{-\frac{\ri \tilde{\Omega}}{2}}f_k(t(\xi)),
\end{align}
where $\Omega$ corresponds to the energy of the daughter particles, $\tilde{\Omega}=\Omega/\mu$ being dimensionless energy, $f_k(t)$ denotes the mode function of the dressed parent particle and $t(\xi)$ implies the original time variable is the function of $\xi=\tanh(\mu t)$. In our model, the mode function is represented by associated Legendre functions. Thanks to the identity~\eqref{QPtrans}, the possible integrations can be represented by
\begin{align}
\tilde{I}_{k,\pm}(\Omega)=\int^{+1}_{-1}d\xi (1-\xi^2)^{-1}(1-\xi)^{\frac{\ri\tilde{\Omega}}{2}}(1+\xi)^{-\frac{\ri \tilde{\Omega}}{2}}P^{\pm\ri r_k}_{\nu}(\xi).\label{IpmOdef}
\end{align}
We use the relation \eqref{PFtrans}, which yields
\begin{align}
    \tilde{I}_{k,\pm}(\Omega)=&\frac{1}{\Gamma(1\mp\ri r_k)}\int^{+1}_{-1}d\xi (1-\xi)^{-1+\frac{\ri}{2}\tilde{\Omega}\mp \frac{\ri}{2}r_k}(1+\xi)^{-1-\frac{\ri }{2}\tilde{\Omega}\pm \frac{\ri}{2}r_k}F\left(-\nu,\nu+1;1\mp\ri r_k;\frac{1-\xi}{2}\right)\nonumber\\
    =&\frac{1}{2\Gamma(1\mp\ri r_k)}\int_0^1 d\zeta\zeta^{-1+\frac{\ri}{2}\rho_k^{\mp}}(1-\zeta)^{-1-\frac{\ri}{2}\rho_k^{\mp}}F\left(-\nu,\nu+1;1\mp\ri r_k;\zeta\right)\nonumber\\
    =&\frac{1}{2\Gamma(1\mp\ri r_k)}\int_0^1 d\zeta\zeta^{-1+\frac{\ri}{2}\rho_k^{\mp}}(1-\zeta)^{-1-\frac{\ri}{2}\rho_k^{\mp}\mp\ri r_k }F\left(\mp\ri r_k+\nu+1,\mp\ri r_k-\nu;1\mp\ri r_k;\zeta\right)
\end{align}
where $\zeta=\frac{1-\xi}{2}$ and $\rho_k^{\mp}\equiv \tilde{\Omega}\mp r_k$. In the last line, we have used the property of the hypergeometric function (see 15.8.1 of~\cite{NIST:DLMF})
\begin{align}
    F(\alpha,\beta;\gamma;z)=(1-z)^{\gamma-\alpha-\beta}F(\gamma-\alpha,\gamma-\beta;\gamma;z).
\end{align}
The above integration cannot be performed as it is, but we are allowed to define it as a limit:
\begin{align}
  \tilde{I}_{k,\pm}(\Omega)=&\lim_{\epsilon\to+0}  \frac{1}{2\Gamma(1\mp\ri r_k)}\int_0^1 d\zeta\zeta^{-1+\epsilon+\frac{\ri}{2}\rho_k^{\mp}}(1-\zeta)^{-1+\epsilon-\frac{\ri}{2}\rho_k^{\pm} }F\left(\mp\ri r_k+\nu+1,\mp\ri r_k-\nu;1\mp\ri r_k;\zeta\right)\nonumber\\
  =&\lim_{\epsilon\to+0}  \frac{1}{2\Gamma(1\mp\ri r_k)}\frac{\Gamma\left(\epsilon+\frac{\ri}{2}\rho_k^{\mp}\right)\Gamma\left(\epsilon-\frac{\ri}{2}\rho_k^{\pm} \right)}{\Gamma\left(2\epsilon\mp\ri r_k\right)}\nonumber\\
  &\qquad \times{}_3F_2\left(\mp\ri r_k+\nu+1,\mp\ri r_k-\nu,\epsilon+\frac{\ri}{2}\rho_k^{\mp};1\mp\ri r_k,2\epsilon\mp\ri r_k;1\right),
\end{align}
where ${}_3F_2(\alpha_1,\alpha_2,\alpha_3;\beta_1,\beta_2;z)$ is a generalized hypergeometric function. The regularization by $\epsilon$ is nothing but multiplication of the adiabatic factor $(4\cosh^2(\mu t))^{-\epsilon}$ to the original integrand such that the integrand vanishes at the asymptotic past and future, namely the interaction is turned off in the asymptotic time region. The second equality follows from the identity (7.512-5 of~\cite{2014776}),
\begin{align}
    \int_0^1d\zeta\zeta^{\rho-1}(1-\zeta)^{\sigma-1}F(\alpha,\beta;\gamma;\zeta)d\zeta=\frac{\Gamma(\rho)\Gamma(\sigma)}{\Gamma(\rho+\sigma)}{}_3F_2(\alpha,\beta,\rho;\gamma,\rho+\sigma;1),
\end{align}
provided that ${\rm Re}\ \rho>0,\ {\rm Re}\ \sigma>0,\ {\rm Re}(\gamma+\sigma-\alpha-\beta)>0$, which can be satisfied by the regularization.
Nevertheless, the limit of $\epsilon\to 0$ is yet ill-defined.\footnote{The generalized hypergeometric function with an argument unity ${}_3F_2(\alpha_1,\alpha_2,\alpha_3;\beta_1,\beta_2;1)$ is convergent if ${\rm Re}(\beta_1+\beta_2-\alpha_1-\alpha_2-\alpha_3)>0$ but it cannot be satisfied in the limit $\epsilon\to 0$.} We further apply another identity in Sec.16.4 of~\cite{NIST:DLMF}
\begin{align}
    {}_3F_2(\alpha_1,\alpha_2,\alpha_3;\beta_1,\beta_2;1)=&\frac{\Gamma(\beta_1)\Gamma(\beta_1+\beta_2-\alpha_1-\alpha_2-\alpha_3)}{\Gamma(\beta_1-\alpha_2)\Gamma(\beta_1+\beta_2-\alpha_1-\alpha_3)}\nonumber\\
    &\qquad\times{}_3F_2(\alpha_2,\beta_2-\alpha_1,\beta_2-\alpha_3;\beta_2,\beta_1+\beta_2-\alpha_1-\alpha_3;1),
\end{align}
for ${\rm Re}(\beta_1+\beta_2-\alpha_1-\alpha_2-\alpha_3)>0$ and ${\rm Re}(\beta_1-\alpha_2)>0$,\footnote{The hypergeometric function ${}_3F_2(\alpha_1,\alpha_2,\alpha_3;\beta_1,\beta_2;z)$ are symmetric in the first three and the second two parameters, and any pair of $(\alpha_i,\beta_j)$ can be chosen instead of a pair $(\alpha_2,\beta_1)$ as long as the conditions are satisfied.} which reads
\begin{align}
   & {}_3F_2\left(\mp\ri r_k+\nu+1,\mp\ri r_k-\nu,\epsilon+\frac{\ri}{2}\rho_k^{\mp};1\mp\ri r_k,2\epsilon\mp\ri r_k;1\right)\nonumber\\
   =&\frac{\Gamma(1\mp\ri r_k)\Gamma\left(\epsilon-\frac{\ri}{2}\rho^{\mp}_k\right)}{\Gamma\left(1+\nu\right)\Gamma\left(\epsilon-\frac{\ri}{2}\rho_k^{\pm}-\nu\right)}\nonumber\\
   &\qquad\times{}_3F_2\left(\mp\ri r_k-\nu,2\epsilon-\nu-1,\epsilon-\frac{\ri}{2}\rho_k^\pm;2\epsilon\mp \ri r_k,\epsilon-\frac{\ri}{2}\rho_k^{\pm}-\nu;1\right).
\end{align}
Using this expression, we are able to take the limit $\epsilon \to 0$ without singularity, and we obtain
\begin{align}
    \tilde{I}_{k,\pm}(\Omega)=&\frac{\pi\Gamma\left(-\frac{\ri}{2}\rho_k^{\pm} \right)}{\rho_k^{\mp}\sinh\left(\frac{\pi}{2}\rho_k^{\mp}\right)\Gamma\left(\mp\ri r_k\right)\Gamma\left(1+\nu\right)\Gamma\left(-\frac{\ri}{2}\rho_k^{\pm}-\nu\right)}\nonumber\\
    &\qquad\times{}_3F_2\left(\mp\ri r_k-\nu,-\nu-1,-\frac{\ri}{2}\rho_k^\pm;\mp \ri r_k,-\frac{\ri}{2}\rho_k^{\pm}-\nu;1\right)
\label{Ipmfinal}
\end{align}
From the definition of the generalized hypergeometric function, we find
\begin{align}
   & {}_3F_2\left(\mp\ri r_k-\nu,-\nu-1,-\frac{\ri}{2}\rho_k^\pm;\mp \ri r_k,-\frac{\ri}{2}\rho_k^{\pm}-\nu;1\right)\nonumber\\
   =& \sum_{N=0}^{\infty}\frac{\Gamma(\mp\ri r_k-\nu+N)\Gamma(-\nu-1+N)\Gamma(-\frac{\ri}{2}\rho_k^\pm+N)\Gamma(\mp \ri r_k)\Gamma(-\frac{\ri}{2}\rho_k^{\pm}-\nu)}{\Gamma(\mp\ri r_k-\nu)\Gamma(-\nu-1)\Gamma(-\frac{\ri}{2}\rho_k^\pm)\Gamma(\mp \ri r_k+N)\Gamma(-\frac{\ri}{2}\rho_k^{\pm}-\nu+N)}\frac{1}{N!}
\end{align}
In the limit $r_k,|\rho^\pm_k|\gg1$, we find 
\begin{align}
    &\sum_{N=0}^{\infty}\frac{\Gamma(\mp\ri r_k-\nu+N)\Gamma(-\nu-1+N)\Gamma(-\frac{\ri}{2}\rho_k^\pm+N)\Gamma(\mp \ri r_k)\Gamma(-\frac{\ri}{2}\rho_k^{\pm}-\nu)}{\Gamma(\mp\ri r_k-\nu)\Gamma(-\nu-1)\Gamma(-\frac{\ri}{2}\rho_k^\pm)\Gamma(\mp \ri r_k+N)\Gamma(-\frac{\ri}{2}\rho_k^{\pm}-\nu+N)}\frac{1}{N!}\nonumber\\
    \to&\sum_{N=0}^{\infty}\frac{\Gamma(-\nu-1-N)}{\Gamma(-\nu-1)N!}=0 \quad \text{ for } \nu>0,
\end{align}
which implies that the hypergeometric function decays as increasing $r_k$ and $|\rho_k^\pm|$. 

Notice also that $\tilde{I}_{k,\pm}(\Omega)$ are regular except the point $\rho_k^-=0$. In physical situations, $\tilde{\Omega}$ is always taken to be positive and therefore $\rho^+_k>0$ but $\rho_k^-$ may vanish. In the limit $\rho_k^-\to 0$, we find
\begin{align}
    \tilde{I}_{k,+}(\Omega)\to&\frac{2}{(\rho_k^{-})^2\Gamma\left(1+\nu\right)\Gamma\left(-\ri\tilde{\Omega}-\nu\right)}\nonumber\\
    &\qquad\times{}_3F_2\left(\ri \rho_k^{-}-\ri\tilde{\Omega}-\nu,-\nu-1,\frac{\ri}{2}\rho_k^{-}-\ri\tilde{\Omega};\ri\rho_k^{-}-\ri \tilde{\Omega},\frac{\ri}{2}\rho_k^{-}-\ri\tilde{\Omega}-\nu;1\right).\label{Ipsing}
\end{align}
We note that from the definition of the generalized hypergeometric function,
\begin{align}
    {}_3F_2\left(-\ri\tilde{\Omega}-\nu,-\nu-1,-\ri\tilde{\Omega};-\ri \tilde{\Omega},-\ri\tilde{\Omega}-\nu;z\right)
    =&\sum_{n=0}^\infty\frac{(-\ri\tilde{\Omega}-\nu)_n(-\nu-1)_n,(-\ri\tilde{\Omega})_n}{(-\ri \tilde{\Omega})_n(-\ri\tilde{\Omega}-\nu)_n}\frac{z^n}{n!}\nonumber\\
    =&\sum_{n=0}^\infty(-\nu-1)_n\frac{z^n}{n!}\nonumber\\
    =&(1-z)^{\nu+1}
\end{align}
where $(a)_n=\Gamma(a+n)/\Gamma(a)$ is Pochhammer symbol. Thus, we find the limit
\begin{align}
    {}_3F_2\left(\ri \rho_k^{-}-\ri\tilde{\Omega}-\nu,-\nu-1,\frac{\ri}{2}\rho_k^{-}-\ri\tilde{\Omega};\ri\rho_k^{-}-\ri \tilde{\Omega},\frac{\ri}{2}\rho_k^{-}-\ri\tilde{\Omega}-\nu;1\right)\to \mathcal{O}(\rho_k^-)
\end{align}
as $\rho_k^-\to 0$. Therefore, $\tilde{I}_{k,+}(\Omega)\propto (\rho^-_k)^{-1}$ as $\rho^-_k\to 0$.  We also find in the limit $\rho_k^-\to 0$
\begin{align}
    \tilde{I}_{k,-}(\Omega)\to&-\frac{\pi^2\Gamma\left(-\frac{\ri}{2}\rho_k^{-} \right)}{2\tilde{\Omega}\sinh\left(\pi\tilde{\Omega}\right)\Gamma\left(\ri \tilde{\Omega}\right)\sin\pi\nu},\label{Imsing}
\end{align}
for $\nu\neq 1,2,\cdots$. Noticing that $\Gamma\left(-\frac{\ri}{2}\rho_k^{-} \right)=\frac{2\ri}{\rho_k^-}\Gamma(1-\frac{\ri}{2}\rho_k^{-})\to \frac{2\ri}{\rho_k^-}$ as $\rho_k^-\to0$, we find that both integral $\tilde{I}_{k,\pm}(\Omega)$ diverges as $(\rho_k^-)^{-1}$ in the limit $\rho_k^-\to0$. We will clarify the origin of the divergence in the main text. For a special case where $\nu$ is a positive integer $\nu=1,2,\cdots$, we should take the limit more carefully and find
\begin{align}
      \tilde{I}_{k,-}(\Omega)\to&\frac{\pi(-1)^{\nu}}{2\tilde{\Omega}\sinh\left(\pi\tilde{\Omega}\right)\Gamma\left(\ri \tilde{\Omega}\right)},\label{Imlimintnu}
\end{align}
which indicates that there appears no singularity in $\rho_k^-\to0$ from $\tilde{I}_{k,-}(\Omega)$.

\bibliographystyle{JHEP}
\bibliography{main.bib}
\end{document}